 \documentclass[twocolumn]{aastex6}

\usepackage{changepage} 
\usepackage{amsmath} 
\usepackage{soul}
\usepackage{lipsum}

\fullcollaborationName{The Friends of AASTeX Collaboration}

\begin{document}


\title{Two Jovian planets around the giant star HD\,202696. \\
A growing population of packed massive planetary pairs around massive stars?}

\author{Trifon Trifonov\altaffilmark{1}, Stephan Stock\altaffilmark{2},
 Thomas Henning\altaffilmark{1}, Sabine Reffert\altaffilmark{2}, Martin K\"{u}rster\altaffilmark{1}, Man Hoi Lee\altaffilmark{3,4}, Bertram Bitsch\altaffilmark{1}, R. Paul Butler\altaffilmark{5}, Steven S. Vogt\altaffilmark{6}}


\affil{$^1$ Max-Planck-Institut f\"{u}r Astronomie, K\"{o}nigstuhl  17, D-69117 Heidelberg, Germany}
\affil{$^2$ Landessternwarte, Zentrum f\"{u}r Astronomie der Universit\"{a}t Heidelberg, K\"{o}nigstuhl 12, D-69117 Heidelberg, Germany}
\affil{$^3$ Department of Earth Sciences, The University of Hong Kong, Pokfulam Road, Hong Kong}
\affil{$^4$ Department of Physics, The University of Hong Kong, Pokfulam Road, Hong Kong}
\affil{$^5$ Department of Terrestrial Magnetism, Carnegie Institution for Science, Washington, DC 20015, USA}
\affil{$^6$ UCO/Lick Observatory, Department of Astronomy and Astrophysics, University of California at Santa Cruz, Santa Cruz, CA 95064, USA}

\email{trifonov@mpia.de}



%
%
%
%




\begin{abstract}

   We present evidence for a new two-planet system around the giant star HD\,202696 (= HIP\,105056, BD\,+26 4118). 
   The discovery is based on public HIRES radial velocity (RV) measurements taken at Keck Observatory
   between 2007 July and 2014 September.
   We estimate a stellar mass of 1.91$^{+0.09}_{-0.14}M_\odot$ for HD\,202696, which is located close to the base of the red giant branch. A two-planet self-consistent 
   dynamical modeling MCMC scheme of the RV data followed by a long-term stability test suggests planetary orbital 
   periods of $P_{\rm b}$ = 517.8$_{-3.9}^{+8.9}$ and $P_{\rm c}$ = 946.6$_{-20.9}^{+20.7}$ days,
   eccentricities of $e_{\rm b}$ = 0.011$_{-0.011}^{+0.078}$ and $e_{\rm c}$ = 0.028$_{-0.012}^{+0.065}$ , 
   and minimum dynamical masses of $m_{\rm b}$ = 2.00$_{-0.10}^{+0.22}$\,$M_{\mathrm{Jup}}$ and 
   $m_{\rm c}$ = 1.86$_{-0.23}^{+0.18}$ $M_{\mathrm{Jup}}$, respectively.
   Our stable MCMC samples are consistent with orbital configurations predominantly 
   in a mean period ratio of 11:6 and its close-by high-order mean-motion commensurabilities with low eccentricities. 
   For the majority of the stable configurations, we find an aligned or anti-aligned 
   apsidal libration (i.e.\ $\Delta\omega$ librating around
   0$^\circ$ or 180$^\circ$), suggesting that the HD\,202696 system is likely dominated by secular 
   perturbations near the high-order 11:6 mean-motion resonance. 
   The HD\,202696 system is yet another Jovian-mass pair around an intermediate-mass star with a  
   period ratio below the 2:1 mean-motion resonance. 
   Therefore, the HD\,202696 system is an important discovery, that  
   may shed light on the primordial disk-planet properties
   needed for giant planets to break the strong 2:1 mean motion resonance and settle in more compact orbits.
  
\end{abstract}




\section{Introduction}

By 2018 November,  647 known multiple-planet systems were reported in the 
literature,\footnote{\url{http://exoplanet.eu}} 144 of which 
were discovered using high-precision radial velocity (RV) measurements.
The RV technique is very successful in determining the orbital 
architectures of multiple extrasolar planetary systems.
In some exceptional cases, $N$-body modeling of precise Doppler data in resonant or near-resonant multiple systems can  
reveal the system's dynamical properties and 
constrain the planets' true dynamical masses and mutual inclinations 
\citep[e.g.][and references therein]{Rivera2010,Tan2013,Trifonov2014,Trifonov2018a,Nelson2016}.
Therefore, multiple-planet systems discovered with the Doppler method are fundamentally important 
in order to understand planet formation and evolution in general.

After publication of the Keck HIRES \citep{Vogt1994} velocity archive 
by \citet{Butler2017}, it became apparent that the early K giant star HD~202696 
most likely hosts another RV multi-planet system.
The HIRES data set comprises 42 precise RVs of HD\,202696 taken between 2007 July and 2014 September,
which we find to be consistent with at least two planets in the Jovian-mass regime
with periods of $\sim$520\,days for the inner planet and $\sim$950\,days for the outer planet. 
The HD\,202696 system is remarkable because of the small orbital separation 
between the two massive planets forming an 
orbital period ratio close to the 9:5  mean-motion resonance (MMR). 
Notable examples of small-separation pairs around intermediate-mass stars are 
24\,Sex, $\eta$~Ceti and HD\,47366 \citep[between 9:5 MMR and 2:1 MMR;][]{Johnson2011,Trifonov2014,Sato2016}; 
HD\,200964 and HD\,5319 \citep[4:3 MMR;][]{Johnson2011,Giguere2015};
and HD\,33844 \citep[5:3 MMR;][]{Wittenmyer2016}.
These systems may indicate that more massive stars tend to form 
more massive planetary systems, a fraction of which  
pass the strong 2:1 MMR during the inward planet migration phase and settle 
at high-order resonant or near-resonant configurations with period ratios $<$ 2:1.
The dynamical characterization of the relatively compact system 
around HD\,202696 may help to enhance our understanding of the formation of 
massive multiple-planet systems around intermediate-mass stars.

This paper is organized as follows. In~Section~\ref{HD202696} we present our estimates of the 
stellar properties of HD\,202696.
Section~\ref{RV_analysis} presents our Keplerian and $N$-body dynamical RV modeling 
analysis, which reveals the planets HD\,202696 b and c.
In Section \ref{dynamical_stability} we present the long-term 
stability analysis of the HD\,202696 system and its dynamical properties.
In Section~\ref{Discussion} we provide a brief discussion of 
the HD\,202696 system in the context of current knowledge of 
compact multiple-planet systems around giant stars.
Finally, Section~\ref{Summary} provides a summary of our results.

\section{Stellar parameter estimates of HD\,202696}
\label{HD202696}

The giant star HD\,202696 (= HIP\,105056, BD\,+26 4118) is a bright $V$~=~8.2 mag star
of spectral class K0III/IV with $B - V=$1.003 mag \citep{ESA}.
The second $Gaia$ data release
\citep[Gaia DR2;][]{Gaia_Collaboration2016, Gaia_Collaboration2018b}
lists a parallax of $\pi$ = 5.28 $\pm$ 0.04 mas, 
an effective temperature $T_{\rm eff}$ = 4951.4 K, a radius estimate
of 6.01 $R_\odot$, and a mean $G$ = 7.96 mag.
\citet{Bailer_Jones} carried out a Bayesian inference using the DR2
parallax information with a prior on distance and estimated a distance of 188.5$_{-1.6}^{+1.6}$ pc.

\begin{table}[htp]

\caption{Stellar parameters of HD202696 and their 1$\sigma$ uncertainties.} 
\label{table:phys_param}    


\centering          
\begin{tabular}{ l l r}     
\hline\hline  \noalign{\vskip 0.5mm}        
  Parameter   &HD202696   &  reference \\  
\hline    \noalign{\vskip 0.5mm}                   
   Spectral type                            & K0III-IV          & [1] \\ 
   $V$ [mag]                           &     $8.23\pm0.01$      & [1] \\ 
   $B-V$ [mag]                           & $1.003\pm0.005$          & [1] \\ 
     E$(B-V)$   [mag]       & 0.08 $\pm$  0.04& [2]     \\    
      A$_V$   [mag]       & 0.25 $\pm$  0.13& [2]     \\
   $\pi$  [mas]                             & 5.28 $\pm$ 0.04  &  [3] \\  
   RV$_{absolute}$   $[$km\,s$^{-1}]$       & --34.45 $\pm$  0.22& [3]     \\    
    Radius    [$R_{\odot}$]   
    & 6.01$_{-0.19}^{+0.33}$  &  [3] \\  
   $T_{\mathrm{eff}}$~[K]                   & 4951$_{-129}^{+78}$ & [3]     \\    
    Distance  [pc]                           & 188.5$_{-1.6}^{+1.6}$   & [4] \\   \noalign{\vskip 0.9mm}  
   
   From Bayesian inference$^\alpha$   &    &  \\    
   \hline    \noalign{\vskip 0.5mm}                   
 
   Mass    [$M_{\odot}$]                    & 1.91$_{-0.14}^{+0.09}$    & This paper\\
   Radius    [$R_{\odot}$]                  & 6.43$_{-0.39}^{+0.41}$    & This paper  \\
   Luminosity    [$L{_\odot}$]              & 23.4$_{-2.0}^{+2.4}$     & This paper \\
   Age    $[$Gyr$]$                         & 1.32$_{-0.25}^{+0.35}$     &  This paper\\  
   $T_{\mathrm{eff}}$~[K]                   & 5040$_{-85}^{+71}$     & This paper \\   
   $\log g~[\mathrm{cm\cdot s}^{-2}]$       & 3.11$_{-0.07}^{+0.07}$    & This paper  \\  
   Evolutionary stage                       & RGB (P$=0.996)$    & This paper  \\ 
    Evolutionary phase$^\beta$              & 8.30$_{-0.11}^{+0.13}$    & This paper  \\  \noalign{\vskip 0.9mm}

   From Spectra$^\gamma$   &    &  \\  
 \hline    \noalign{\vskip 0.5mm}                   
   $T_{\mathrm{eff}}$~[K]                   & 4988 $\pm$ 57     & This paper \\
   $\log g~[\mathrm{cm\cdot s}^{-2}]$       & 3.24 $\pm$ 0.15    & This paper \\   
   {}[Fe/H]                                 & 0.02 $\pm$ 0.04   & This paper  \\
   $v\cdot\sin(i)$~$[$km\,s$^{-1}]$              & 2.0  $\pm$ 0.8    & This paper   \\                                          
   RV$_{absolute}$   $[$km\,s$^{-1}]$       & --33.17 $\pm$  0.05& This paper     \\ 
   
\hline\hline \noalign{\vskip 0.5mm}   

\end{tabular}


%


\tablecomments{\small $\alpha$ Following \citet{Stock2018}, $\beta$ A value of $8.0$ marks the RG base while $9.0$ marks the RGB bump \citep{Bressan2012}, $\gamma$ from HIRES 
spectra using $CERES$ \citep{Brahm2017a} and $ZASPE$ \citep{Brahm2017b},  
[1] \citet{ESA}, [2] \citet{Gontcharov2018}, [3] \citet{Gaia_Collaboration2016, Gaia_Collaboration2018b} , [4] \citet{Bailer_Jones}  
 }

\end{table}

\begin{figure}
\resizebox{\hsize}{!}{\includegraphics{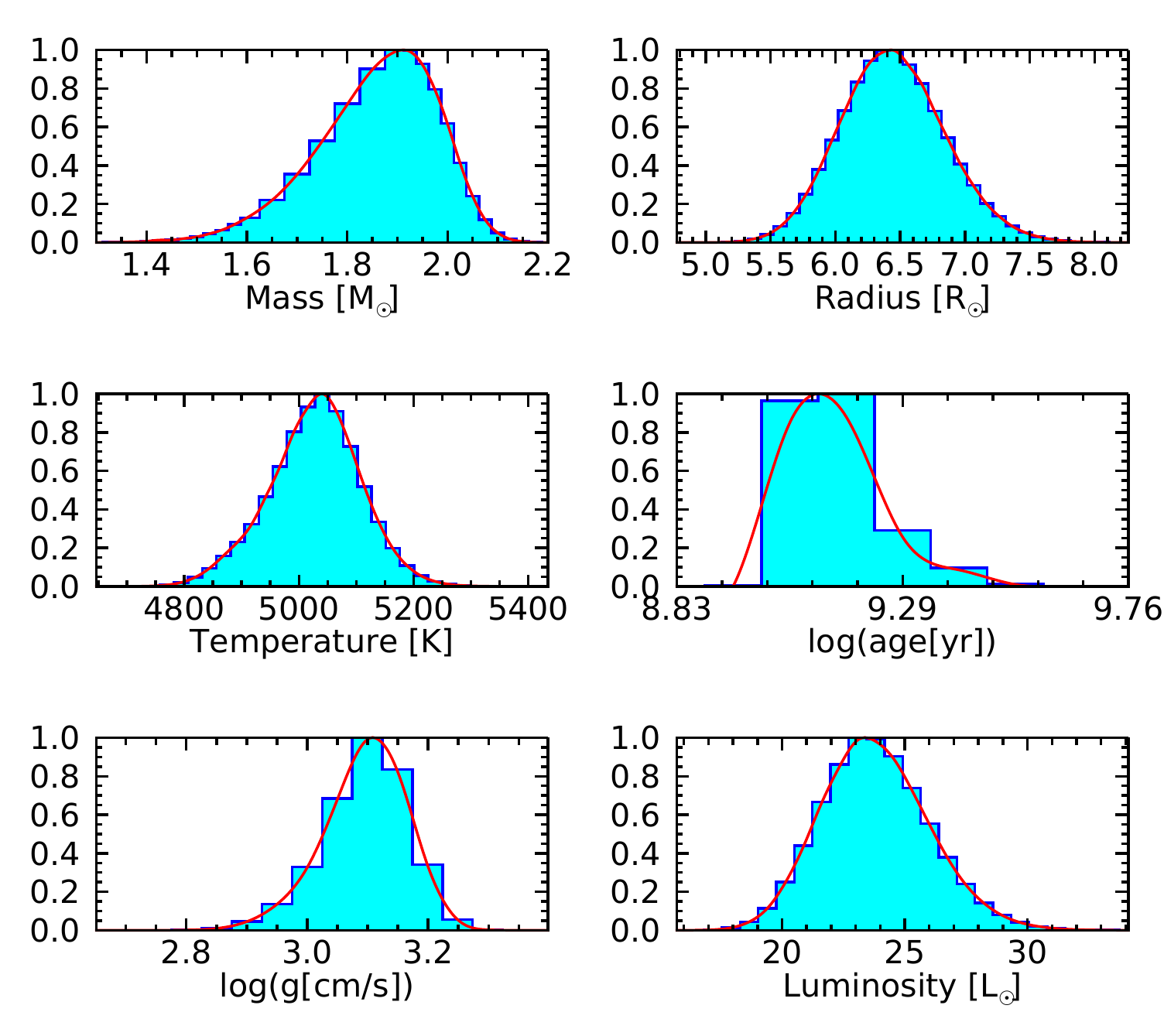}}
\caption{Bayesian stellar parameters estimated using the $Gaia$ DR2 parallax, $Hipparcos B$ and $V$ photometry, and the spectroscopic metallicity.}
\label{rgb} 
\end{figure}

For the dynamical analysis of the HD\,202696 system, we need to incorporate the stellar 
parameters of the host star, most importantly its stellar mass. This is achieved by applying a 
slightly modified version of the methodology and Bayesian inference scheme used in \cite{Stock2018}, which was 
particularly optimized for subgiant and giant stars such as HD\,202696 and is capable of determining
the post-main-sequence evolutionary stage.
The method was shown to provide robust estimates of stellar parameters such as mass $M$, 
radius $R$, age $\tau$, effective temperature $T_{\mathrm{eff}}$, surface gravity $\log g$, and 
luminosity $L$ in agreement with other methods, such as asteroseismology, spectroscopy, and interferometry. 
The method uses the trigonometric parallax, a metallicity estimate, and photometry 
in at least two different bands as input parameters. 
We decided to use the $Hipparcos$ $V$ and $B$ photometry as in \cite{Stock2018} instead of the available 
high-precision $Gaia$ DR2 photometry because it was shown that results based on this choice of photometry 
are in good agreement with literature values. While $Gaia$ DR2 photometry is more precise, 
it has yet to be tested whether comparisons with stellar evolutionary models do not suffer from systematics, 
which is especially important regarding the small photometric uncertainties.
\cite{Stock2018} adopted the stellar models based on 
the PAdova and TRieste Stellar Evolution Code \citep[PARSEC;][]{Bressan2012} and used bolometric 
corrections by \cite{Worthey}. The Bayesian inference is normally carried out in the plane of 
the astrometric-HR diagram \citep{Arenou1999} which has a color as the abscissa and the 
astrometry-based luminosity (ABL) in a specific photometric band $\lambda$ as the ordinate. 
The ABL is a quantity that is linear in the trigonometric parallax, allowing for unbiased 
comparisons of stellar positions to evolutionary models if the parallax error dominates. 
However, HD\,202696 is affected by a significant amount of extinction, which in addition is 
very uncertain. \cite{Gontcharov2018} estimated a reddening of $E(B-V)=0.08$ mag and an 
extinction of $A_V=0.25$ mag in the V band for HD\,202696. They also stated that their reddening 
estimates are better than 0.04 mag, a value that we conservatively adopt as the formal reddening error for HD\,202696.
 We used the python SED fitting tool by \cite{Robitaille} together with the model atmospheres of
 \cite{Castelli2004} to fit the available broadband photometry of HD\,202696.
We derived an extinction of $A_{V}=0.144^{+0.280}_{-0.012}$ mag, consistent with the value and uncertainties
by \cite{Gontcharov2018}. However, due to the sparse sampling of the model atmospheres, the errors of the SED 
fitting are probably underestimated and should be regarded with caution. For the analysis of the stellar 
parameters in this paper, we used the extinction by \cite{Gontcharov2018}. As the uncertainty of the 
extinction is the dominating factor in the determination of the stellar parameters of HD\,202696,
it is justified to use the absolute magnitude of the star instead of the ABL for the determination of stellar
parameters  and include the uncertainties of the extinction. We use the Bayesian distance estimate by \cite{Bailer_Jones} 
for the determination of the absolute magnitude from the distance modulus. This decreases biases that arise when determining 
the absolute magnitude from the trigonometric parallax, which is why the ABL should normally be the preferred quantity.

 \begin{figure}
\resizebox{\hsize}{!}{\includegraphics{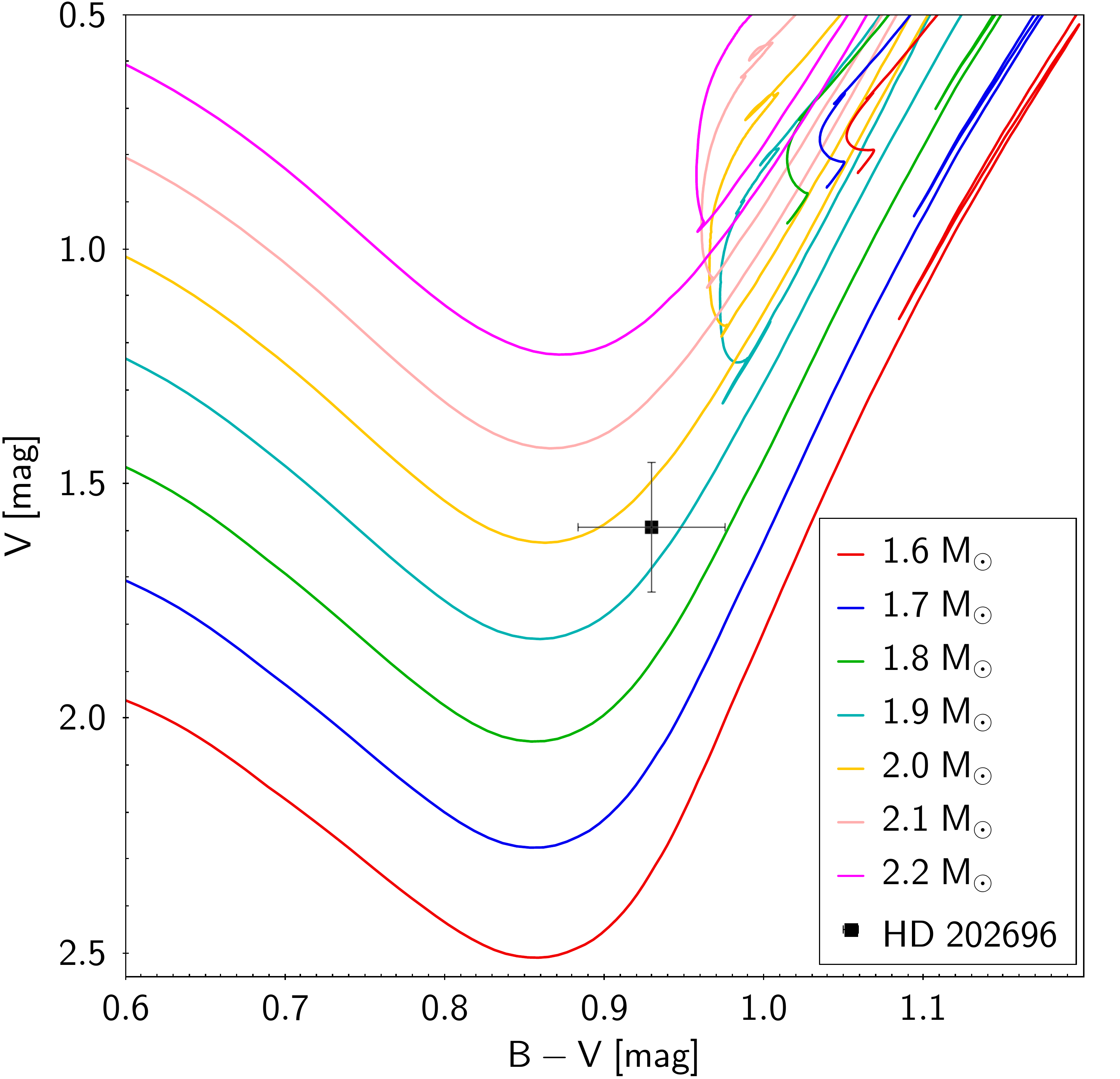}}
\caption{Padova evolutionary tracks of metallicity $Z=0.015875$ ([Fe/H]$\approx0.02$)
with various initial stellar masses from $1.6$ (red line) 
to $2.2$ (pink solid line) $M_\odot$ in steps of $0.1\;$M$_\odot$ in the CMD.
The position of HD\;202696 is highlighted as the black point. }
\label{evtrack} 
\end{figure}

We need to determine the spectroscopic metallicity of HD\,202696 in order to
derive the stellar mass. We derive spectroscopic stellar parameters such as $T_{\rm eff}$, 
surface gravity log $g$, metallicity [Fe/H], and $v\cdot\sin(i)$ 
by using the available HIRES spectral observations of HD\,202696.
As the spectra obtained at Keck Observatory have severe iodine (I$_{\rm 2}$) line 
contamination imposed in order to measure RVs using the I$_{\rm 2}$ 
method \citep[][]{Marcy2,Valenti,Butler1996}, we use the single I$_{\rm 2}$-free stellar template 
spectrum\footnote{Downloaded from \url{http://nexsci.caltech.edu/archives/koa/}} 
required for the RV modeling process \citep[see][]{Butler1996} to determine our 
spectroscopic parameters. We reduce the I$_{\rm 2}$-free spectrum, with a 
signal-to-noise ratio of approximately 120, using the CERES pipeline \citep{Brahm2017a} and then 
analyze it by using ZASPE \citep{Brahm2017b}. Table~\ref{table:phys_param} provides our
spectral stellar parameter estimates of HD~202696.

Using the spectroscopic metallicity estimate, the $B$ and $V$ photometry of the $Hipparcos$ 
catalog and the Bayesian distance estimate of \citet{Bailer_Jones}, together with their uncertainties,
we apply the slightly modified Bayesian inference method of \citet{Stock2018} 
and determine that HD\,202696 is, with a probability of 99.6\%, an early red giant branch star,
right at the beginning of its ascent. We neglect the small probability of the star belonging to the horizontal branch. We estimate
a stellar mass of $M$~=~1.91$_{-0.14}^{+0.09}$~$M_{\odot}$,
a luminosity of $L$~=~23.4$_{-2.4}^{+2.0}$~$L_{\odot}$, and an 
age of $1.32_{-0.25}^{+0.35}$ Gyr. The derived effective temperature of $T_{\mathrm{eff}}=5040_{-85}^{+71}$~K is 
within 1\,$\sigma$ of our spectroscopic estimate (4988 $\pm$ 57~K) and the estimate by $Gaia$ DR2 (4951$_{-129}^{+78}$~K). 
The same is true for the surface gravity of $\log g=3.11_{-0.07}^{+0.07}$ dex when compared to our 
spectroscopic estimate (3.24 $\pm$ 0.15 dex) and the radius $R$~=~6.43$_{-0.39}^{+0.41}$~$R_\odot$ 
when compared to the $Gaia$ DR2 estimate (6.01$_{-0.19}^{+0.33}$ $R_\odot$). 
Given the Bayesian stellar radius and the $v\cdot\sin(i)$ = 2.0 $\pm$ 0.8 km\,s$^{-1}$
estimated from the spectra, we calculate a stellar rotational period of $P_{\rm rot} = 162.7 \pm 65.9$ days, which is typical for an early evolved star such as HD\,202696.

The good agreement between the Bayesian stellar parameters and the other estimates strengthens 
our confidence in the adopted extinction of the star, as the extinction is significant and our 
results rather sensitive to the precise extinction value. 
Neglecting extinction, the resulting mass would have been $\sim 1.4\,M_{\odot}$, 
considerably smaller than before, and the stellar parameters would not have 
been consistent with constraints provided by spectroscopy and by Gaia DR2 data. 
However, the fact that extinction and reddening are consistent with zero at the 2\,$\sigma$ 
level leads to a relatively large uncertainty of the stellar mass, especially toward lower masses. 
The posterior distribution for each of the stellar parameters determined from the Bayesian inference is shown in
Fig.~\ref{rgb}, while Fig.~\ref{evtrack} shows the Padova evolutionary tracks in the color-magnitude diagram (CMD) for various initial stellar masses and the CMD position of HD\,202696. 
The Bayesian stellar parameters, which were determined from the maximum of the distribution, along with their $1\sigma$ error estimates are summarized in Table~\ref{table:phys_param}.

%



\begin{figure}[btp]

 \begin{center}$
\begin{array}{ccc}

\includegraphics[width=9cm]{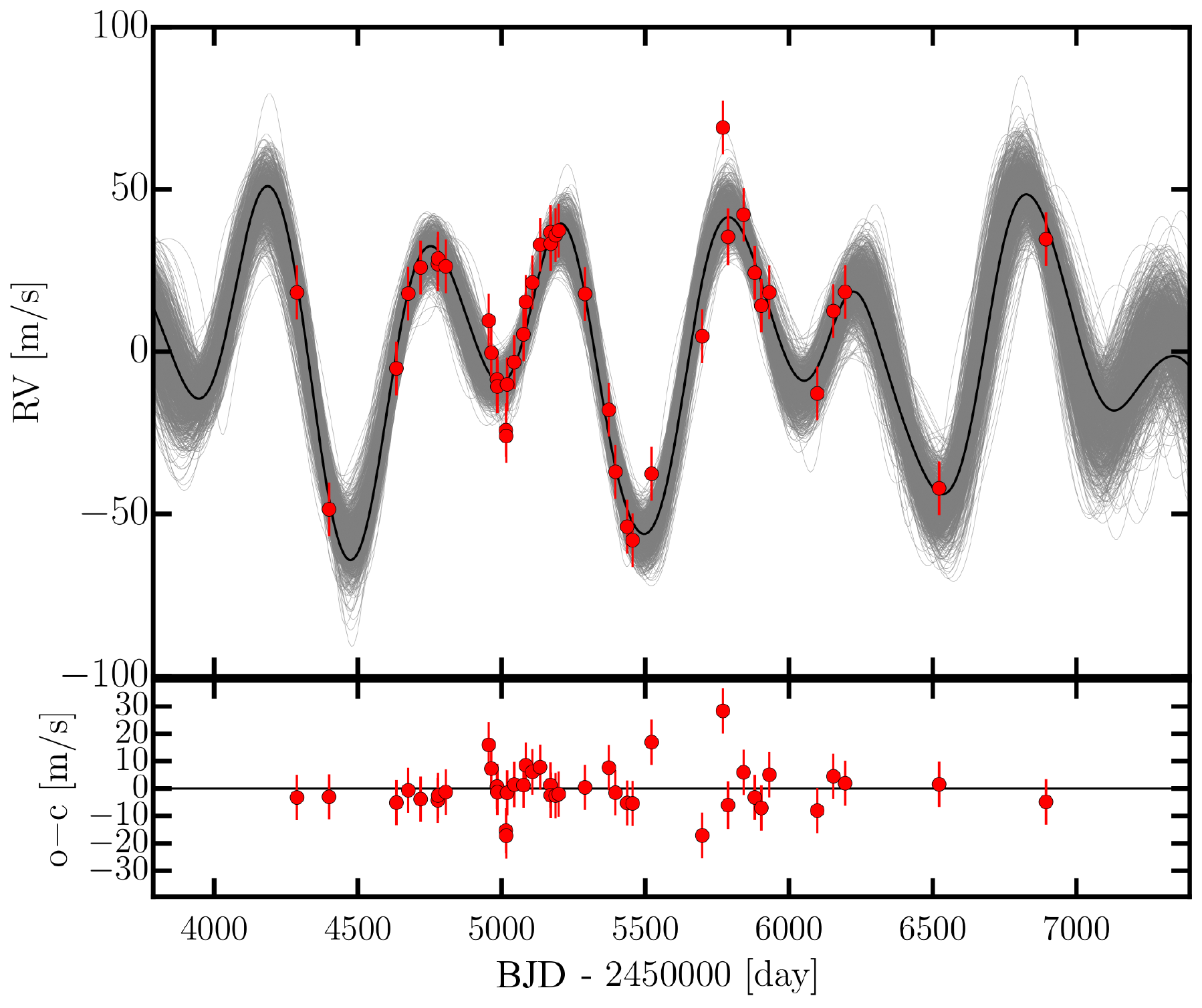}

\end{array} $

\end{center}

\caption{The top panel shows the available precise RV data of HD\,202696 measured 
with HIRES between 2007 July  and 2014 September.
The solid curve is the best two-planet Keplerian model fitted to the data, while 
the shaded area is composed of 1000 randomly chosen confident 
fits from an MCMC test constructed to study the parameter distribution. 
The bottom panel shows the best-fit residuals. After removing the two Keplerian 
components of the RV signal with 
semi-amplitudes of $\sim$ 32 and 28 m\,s$^{-1}$ the RV scatter has an rms of 8.3 m\,s$^{-1}$, which is mostly consistent 
with the expected short-period stellar jitter for this target.
}
\label{diff_fit} 

\end{figure}

\begin{figure}[btp]
 
 \begin{center}$
\begin{array}{ccc}

\includegraphics[width=9cm]{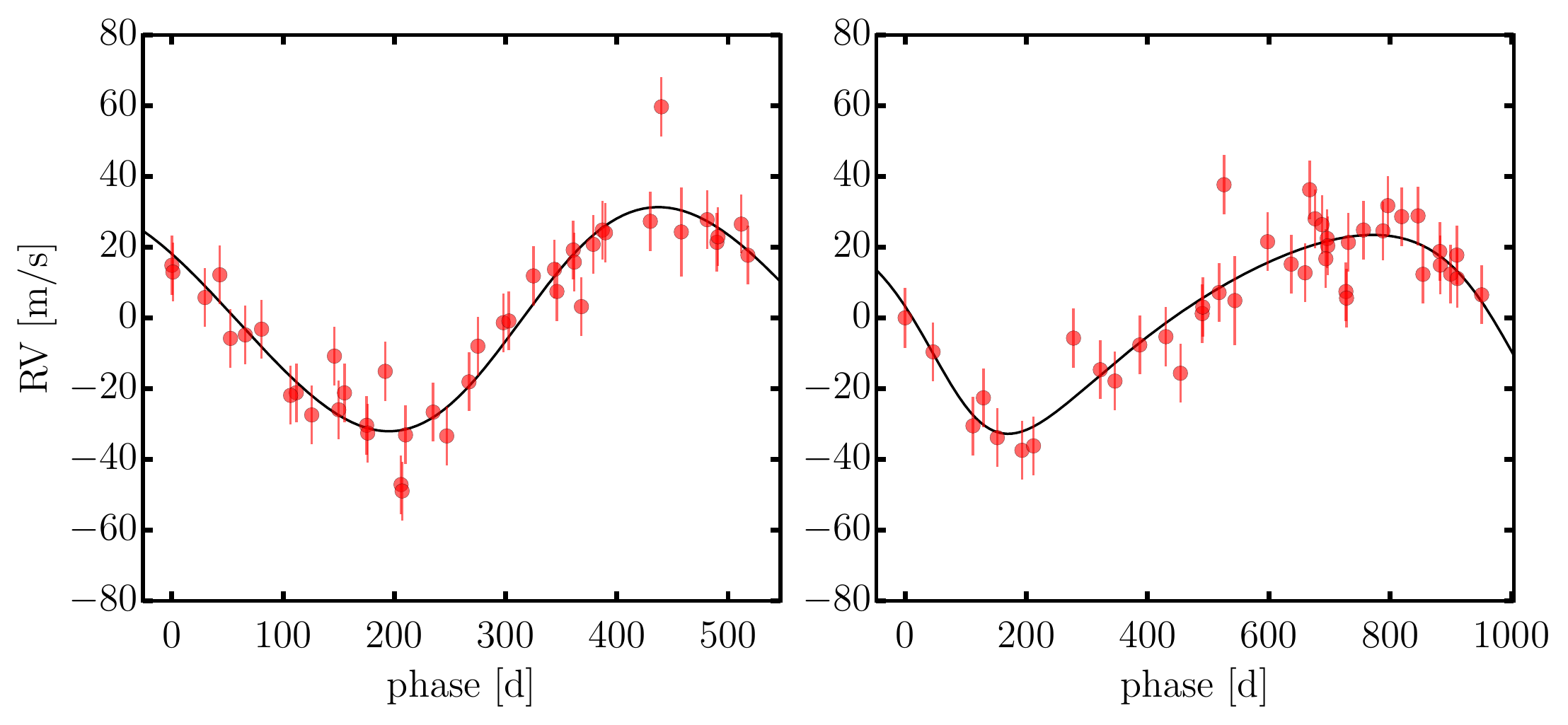}

\end{array} $

\end{center}

\caption{Two Keplerian signals with periods of 521.0 days and 956.1 days, respectively, shown in a phase-folded representation.}
\label{phase_plot} 

\end{figure}

\section{Analysis of RV data set}
\label{RV_analysis} 

\subsection{Period search}
\label{gls_analysis}

The HIRES RV time series of HD\,202696 consists of 42 precise Doppler measurements taken
between 2007 July and 2014 September. With a median RV precision of 1.37 m\,s$^{-1}$
and an end-to-end amplitude of 127\,m\,s$^{-1}$, these data clearly show systematic motions, 
resembling two Keplerians caused by two massive orbiting planets.
Figure~\ref{diff_fit} shows the precise RV data of HD\,202696 together with our best two-planet Keplerian model
and its uncertainties, while Fig.~\ref{phase_plot} shows the two fairly well-sampled 
Keplerian signals phase-folded at their best-fit periods (see Section~\ref{kep_dyn_analysis} for a quantitative analysis of the data).
Note that the data uncertainties displayed in Figures~\ref{diff_fit} and 
\ref{phase_plot} already contain the best-fit RV jitter 
estimate of $\sim $8 m\,s$^{-1}$, which was automatically added in quadrature to the error budget during the fitting.
This RV jitter estimate is typical for early K giants such as HD~202696 and is likely due to solar-like oscillations, 
which have periods of less than a day and are thus not resolved in the 
long-term RV time series. The scaling relations from \cite{Kjeldsen2011} 
predict 3.7\,m\,s$^{-1}$ RV jitter for HD\,202969, while the typical RV jitter value for 
a star with a $B-V$ color of 1.003 mag in the Lick sample is 10--20\,m\,s$^{-1}$ \citep[see Fig.\,3 in][]{Trifonov2014}. 
Since the stars in the Lick sample are, on average, slightly more evolved than HD\,202696, 
it might be reasonable to expect HD\,202696 to have slightly smaller RV jitter than most other stars in this sample.

It is worth noting that with the release of the HIRES database, \citet{Butler2017} 
reported the presence of at least three periods in
the Keck data of HD\,202696. They identified two significant signals  
with periods of 522.3 $\pm$ 6.4 and 946.0 $\pm$ 19.0\,days, classified as planetary 
``candidates'', and an additional period of 214.2 $\pm$ 4.8\,days which required further confirmation.

\begin{figure}[btp]
 
\begin{center}$
\begin{array}{cc} 
\includegraphics[width=9cm]{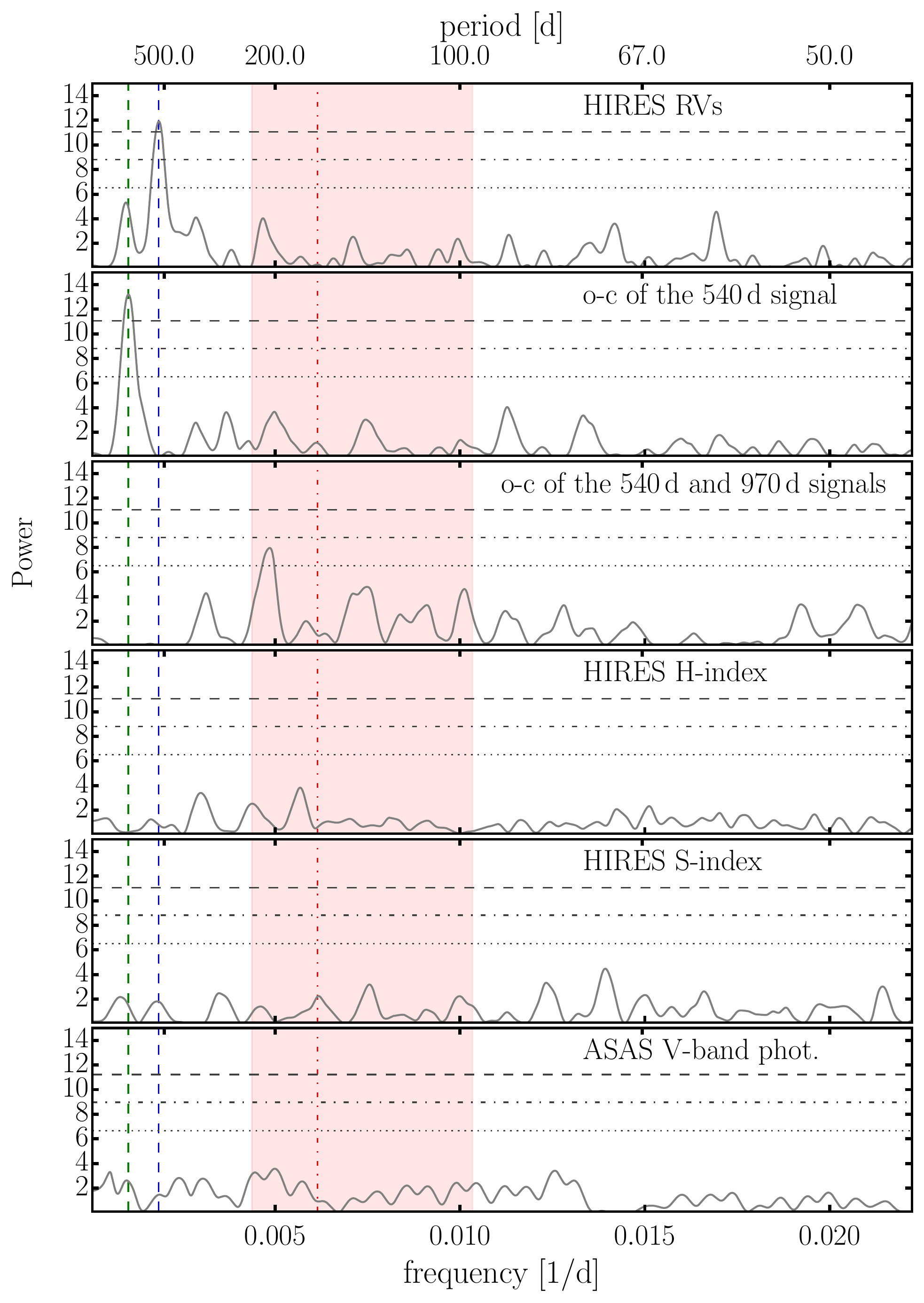} 
\put(-40,327){\tiny p = 0.1\%} 
\put(-40,319){\tiny p = 1\%} 
\put(-40,312){\tiny p = 10\%} 
\end{array}$
\end{center}

\caption{The GLS periodograms of the available HIRES RV time series reveal two significant signals around $\sim$540 and $\sim$970 days,
marked with blue and green dashed lines, respectively. The possible stellar rotation frequencies are within the red shaded area,
the red line denoting the most likely stellar rotation frequency (see Sec.~\ref{HD202696}).
No evident periodicity with an FAP $<$ 0.1\% is detected after the prewhitening  (i.e.\ consecutive  subtraction of each signal from the data.)
These two signals do not have counterparts in the S- and H-index activity indicators of the HIRES data, 
nor do they fall into the range of possible stellar rotational frequencies.
The mean ASAS $V$-band photometry time series of HD\,202696 is also clean of significant periodicities at these periods.
}
\label{gls} 

\end{figure}

As an independent test, we derive the generalized Lomb-Scargle periodogram \citep[GLS;][]{Zechmeister2009} 
of the HIRES RV data, as well as that of the S- and H-index activity indicators (Fig.~\ref{gls}) also provided by \citet{Butler2017}.
For reference, in all panels of Fig.~\ref{gls} the range of possible stellar rotation frequencies of HD\,202696 is indicated by the red shaded area, 
with the red dashed line at the most likely value (1/$P_{\rm rot.}$ = 0.00615$_{-0.00177}^{+0.00418}$\,day$^{-1}$, see Section~\ref{HD202696}). 
The horizontal dotted, dot-dashed, and dashed lines 
correspond to false-alarm probability (FAP) levels of 10\%, 1\%, and 0.1\%, 
respectively. We use an FAP $<$0.1\% as our significance threshold. 
The top panel of Fig.~\ref{gls} reveals a dominant signal around 540\,days (blue dashed line) in the HIRES RVs, which is significant. 
The second panel of Fig.~\ref{gls} shows the periodogram of the RV data
after the  $\sim$ 540\,days signal has been subtracted; it indicates
a second significant signal with a period of $\sim$ 970\,days (green dashed line). The
middle panel of Fig.~\ref{gls} shows that no period with an FAP $<$ 0.1\% can be 
detected after the two most significant signals have been consecutively 
subtracted from the data.

The $\sim$214\,d signal previously reported by \citet{Butler2017} is also 
visible in the residuals of the data, but with 1\% $<$ FAP $<$ 10\%, which we consider insignificant.
Moreover, the 214\,day signal falls within the red shaded area of possible 
stellar rotation frequencies and thus is not considered further in our analysis.
No significant periodicities can be identified in the two panels of Fig.~\ref{gls}, 
which show the periodograms of the H- and the S-index activity indicators, respectively.

Additionally, we inspected the $V$-band measurements of HD\,202696 from the 
All-Sky Automated Survey \citep[ASAS;][]{Pojmanski2005}. 
We find only 48 high-quality ASAS $V$-band measurements 
(graded A and B with $\overline{\sigma_{\rm V}}$ = 0.037 mag.), 
mostly taken between  2003 May and  2007 October, i.e. 
prior to the HIRES Doppler observations of HD\,202696.
Thus, the sparse ASAS photometry is only relevant to check for evident intrinsic photometric 
variability, which could be associated with the RV signals.
No significant periodicity is observed in any of the five available ASAS apertures.
The bottom panel of Fig.~\ref{gls} shows the GLS periodogram of the averaged ASAS photometry measurements, 
indicating that HD\,202696 is most likely a photometrically stable star.

The two significant signals at $\sim$ 540 and $\sim$ 970\,days are consistent 
with those reported by \citet{Butler2017} as planet candidates. We find that 
these signals are not consistent with the range of possible stellar rotation
frequencies of HD\,202696. Also, the S- and H-index activity indicators of 
the HIRES data and the ASAS photometry do  not show any significant 
periodic structure or correlation that could be associated with the RV signals. 
Thus, we consider the $\sim$ 540  and $\sim$ 970\,day signals as 
robust planetary detections.

\subsection{Orbital parameter estimates}
\label{kep_dyn_analysis}

In order to determine the orbital parameters of the HD\,202696 b and c planetary companions,
we adopt a maximum-likelihood estimator (MLE) scheme 
coupled with a Nelder-Mead algorithm \citep[also known as a downhill simplex algorithm; see][]{NelderMead,Press}.
In our scheme, we minimize the negative logarithm of the model's likelihood function ($-\ln \mathcal{L}$),  
while we  simultaneously optimize the planetary RV semi-amplitudes $K_{\rm b,c}$, periods $P_{\rm b,c}$, 
eccentricities $e_{\rm b,c}$, arguments of periastron $\omega_{\rm b,c}$, 
mean anomalies $M_{\rm b,c}$ and the RV zero-point offset. 
 Prior information on the period, phase, and amplitude for each of the two-planet candidates 
is taken from the GLS periodogram test (see Fig.~\ref{gls}), which provides good starting parameters for the MLE algorithm.
All parameters are derived for a uniform reference epoch of 2,450,000 (BJD). 
Following \citet{Baluev2009}, we also include the RV 
jitter\footnote{The RV jitter is most likely due to a 
combination of astrophysical stellar noise and possible instrument 
and data reduction systematics.} as an additional parameter in the modeling.

\begin{table}[ht]

\centering   
\caption{Best Keplerian and dynamical fits for the HD~202696 system.}   
\label{table:orb_par_stable}      

\begin{tabular}{lrrr}     

\hline\hline  \noalign{\vskip 0.7mm}      

\multicolumn{4}{c}{Two-planet Keplerian Model}           \\

\hline \noalign{\vskip 0.7mm}  

Parameter &\hspace{0.0 mm} & HD~202696 b & HD~202696 c  \\

\hline\noalign{\vskip 0.5mm}

Semi-amplitude $K$  [m\,s$^{-1}$]                        & &  31.7$_{-2.1}^{+4.8}$         &  28.1$_{-5.3}^{+1.1}$        \\
Period $P$ [days]   			          & &  521.0$_{-7.3}^{+6.2}$       &  956.1$_{-29.9}^{+22.4}$   \\  
Eccentricity $e$                                       & &  ~~0.056$_{-0.040}^{+0.066}$ &  ~~0.261$_{-0.202}^{+0.060}$ \\
Arg. of periastron $\omega$ [deg]                            &  & ~~259.2$_{-89.2}^{+106.0}$  &  129.1$_{-33.6}^{+142.7}$         \\   
Mean anomaly $M_0$ [deg]                               & &  ~~69.7$_{-115.8}^{+85.8}$   &  152.0$_{-74.3}^{+112.8}$         \\  \noalign{\vskip 0.9mm} 
Mean longitude $\lambda$ [deg]                           & &  ~~328.8$_{-44.9}^{+50.4}$  &  281.1$_{-58.9}^{+50.5}$    \\  
Semi-major axis $a$ [au]                                  & &  1.573$_{-0.015}^{+0.012}$                &  2.357$_{-0.049}^{+0.037}$                     \\  
Minimum mass $m \sin i$ [$M_{\mathrm{Jup}}$]           & &  1.930$_{-0.132}^{+0.285}$                &  2.028$_{-0.354}^{+0.073}$                    \\ \noalign{\vskip 0.9mm}
RV offset~[m\,s$^{-1}$]        & &  \multicolumn{2}{c}{--15.32$_{-1.15}^{+2.59}$}  \\  \noalign{\vskip 0.9mm} 
RV jitter [m\,s$^{-1}$]                      & &  \multicolumn{2}{c}{8.20$_{-0.46}^{+3.07}$}   \\
$-\ln\mathcal{L}$                       
& & \multicolumn{2}{c}{148.799}  
\\
\noalign{\vskip 0.5mm}

\hline\hline\noalign{\vskip 1.2mm} 

\multicolumn{4}{c}{Two-planet Dynamical Model}           \\

\hline \noalign{\vskip 0.7mm}  

Parameter &  & HD~202696 b & HD~202696 c  \\
\hline\noalign{\vskip 0.5mm}

Semi-amplitude $K$ [m\,s$^{-1}$]                        &  & 32.5$_{-1.6}^{+3.1}$          &  26.6$_{-3.5}^{+1.1}$        \\
Period $P$ [days]   		        	          &  & 528.2$_{-12.5}^{+1.8}$        &  958.0$_{-26.4}^{+2.2}$      \\  
Eccentricity $e$                                       &  & ~~0.026$_{-0.006}^{+0.100}$   &  ~~0.265$_{-0.229}^{+0.005}$ \\
Arg. of periastron $\omega$ [deg]                            &  & ~~289.7$_{-91.8}^{+35.1}$    &  ~~140.0$_{-44.9}^{+88.7}$       \\   
Mean anomaly $M_0$ [deg]                               & &  ~~70.2$_{-40.8}^{+67.9}$    &  ~~155.5$_{-52.2}^{+70.41}$    \\  \noalign{\vskip 0.9mm} 
Mean longitude $\lambda$ [deg]                           & &  ~~359.9$_{-63.0}^{+8.2}$    &  259.5$_{-68.5}^{+1.0}$    \\ 
Semi-major axis $a$ [au]                                  &  & 1.587$_{-0.025}^{+0.004}$     &  2.360$_{-0.040}^{+0.003}$                     \\  
Min. dyn. mass $m$ [$M_{\mathrm{Jup}}$]                  & &  1.991$_{-0.011}^{+0.174}$     &  1.917$_{-0.212}^{+0.116}$                    \\ \noalign{\vskip 0.9mm}

Inclination $i$ [deg]                                  & & 90.0 (fixed)     & 90.0 (fixed)                \\ 
Node $\Delta \Omega$ [deg]                             & &  \multicolumn{2}{c}{0.0 (fixed) }                    \\  

RV offset [m\,s$^{-1}$]                       &  & \multicolumn{2}{c}{--15.12$_{-1.44}^{+1.95}$}  \\ \noalign{\vskip 0.9mm} 
RV jitter [m\,s$^{-1}$]                      &  & \multicolumn{2}{c}{8.31$_{-0.47}^{+3.08}$}   \\
 $-\ln\mathcal{L}$                       &   &  \multicolumn{2}{c}{148.973} \\                                     
\noalign{\vskip 0.5mm}

\hline\hline\noalign{\vskip 1.2mm}  

%
%
%

\end{tabular}

%

 \tablecomments{\small The parameters are valid for BJD = 2,450,000 and are derived using our stellar mass estimate of $M$ = 1.91 $M_\odot$.
 The uncertainties of $a_{\rm b,c}$, $m_{\rm b,c} \sin i$ (Keplerian model), and $m_{\rm b,c}$ (dynamical model) do not take into account the small uncertainty of the stellar mass.
}

\end{table}

We apply two models to the RV data: a standard unperturbed two-planet Keplerian 
model and a more accurate self-consistent $N$-body dynamical model, which models
the gravitational interactions between the planets by integrating the equations of motion using the 
Gragg-Bulirsch-Stoer integration method \citep[][]{Press}.
We cross-check the obtained best-fit results for consistency with 
the Systemic~Console 2.2 package \citep{Meschiari2009} 
and find excellent agreement between our MLE
code\footnote{the source code and GUI of our tools can be found in \url{https://github.com/3fon3fonov/trifon} (Trifonov et al. 2018 in preparation)} and Systemic.

For parameter distribution analysis and uncertainty estimates, we couple our MLE fitting algorithm with a 
Markov chain Monte Carlo (MCMC) sampling scheme using the emcee sampler \citep{emcee}. 
For all parameters, we adopt flat priors (i.e.\ equal probability of occurrence) and we run emcee 
from the best fit obtained by the MLE. 
We select the 68.3\% confidence levels of the posterior MCMC parameter distribution as 1$\sigma$ parameter uncertainties.

The best-fit parameters of our two-planet Keplerian and dynamical models, together
with their 1$\sigma$ MCMC uncertainties, are tabulated in Table~\ref{table:orb_par_stable}.
We first fit the Keplerian model and as a next step, we adopt the best two-planet 
Keplerian parameters as an input for the more sophisticated dynamical model. 
For consistency with the unperturbed Keplerian frame and in order to work with
minimum dynamical masses,
we assume an edge-on and coplanar configuration for the HD\,202696 system
(i.e.\ $i_{\rm b,c}$ = 90$^\circ$ and $\Delta\Omega$ = 0$^\circ$).
We find that the models are practically equivalent, suggesting 
a pair of planets with equal masses of the order of $\sim$2.0\,$M_{\rm Jup}$
and a period ratio of $P_{\rm rat.} \approx 1.83$, near the 11:6 MMR 
(for details, see Table~\ref{table:orb_par_stable}).

\begin{figure*}[btp]
\begin{center}$
\begin{array}{cc}

\includegraphics[width=17.2cm]{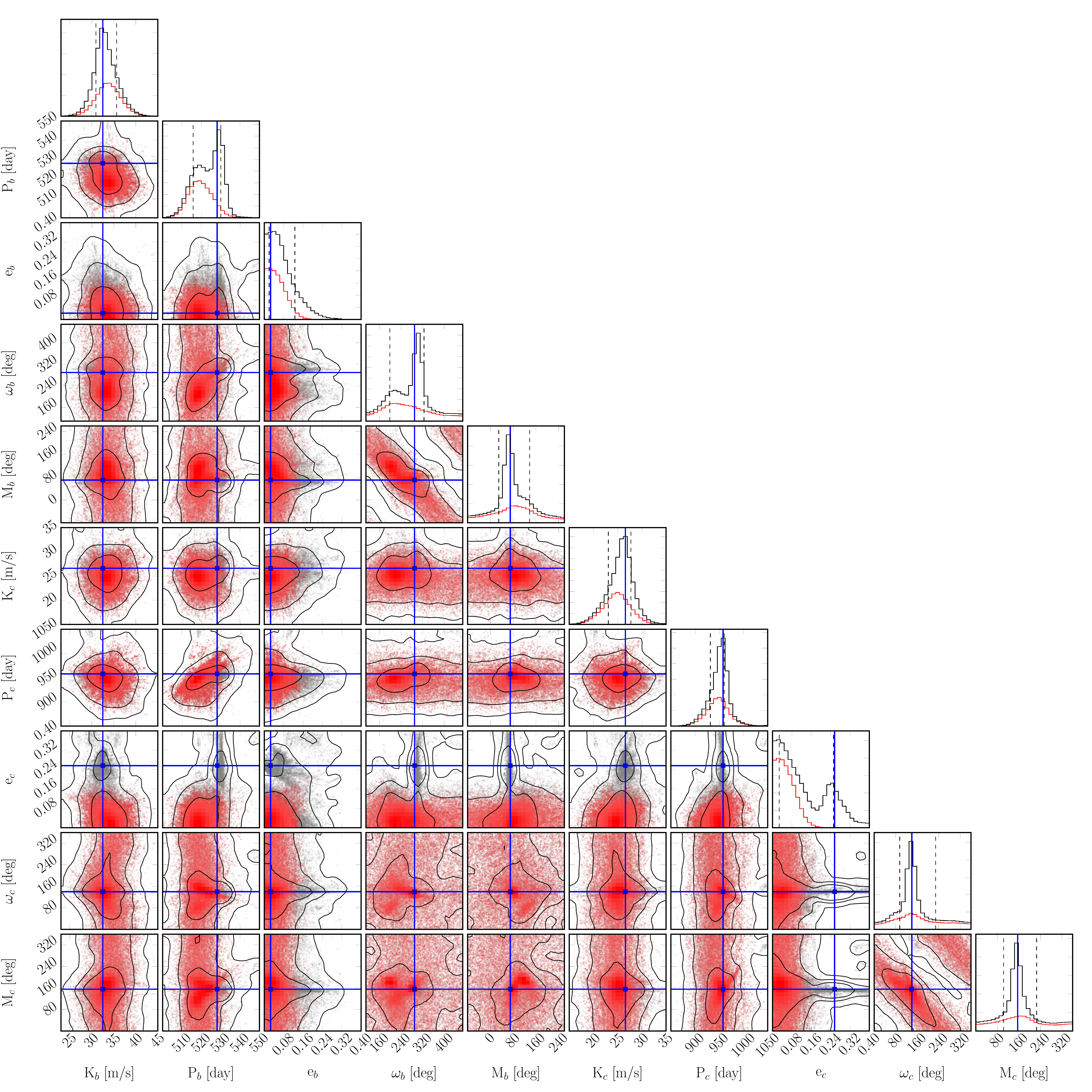}

\end{array} $
\end{center}

 \caption{The MCMC distribution of orbital parameters consistent with the HIRES RV data of HD\,202696 
assuming an edge-on coplanar two-planet system fitted with a self-consistent dynamical model.
The position of the best dynamical fit from Table~\ref{table:orb_par_stable} is marked with blue lines.  
The black contours on the 2D panels represent the 1$\sigma$, 2$\sigma$ and 3$\sigma$ confidence levels of the overall MCMC samples,
while on the 1D histograms, the dashed lines mark the 16th and the 84th percentiles of the the overall MCMC samples.
The samples that are stable for at least 1 Myr are red 
and represent $\sim$46\% of all samples. The stable samples are  
mostly within the 1$\sigma$ confidence region from the best fit. See text for details. 
}  
\label{mcmc} 
\end{figure*}

\begin{figure}[]
\begin{center}$
\begin{array}{ccc}

\includegraphics[width=9cm]{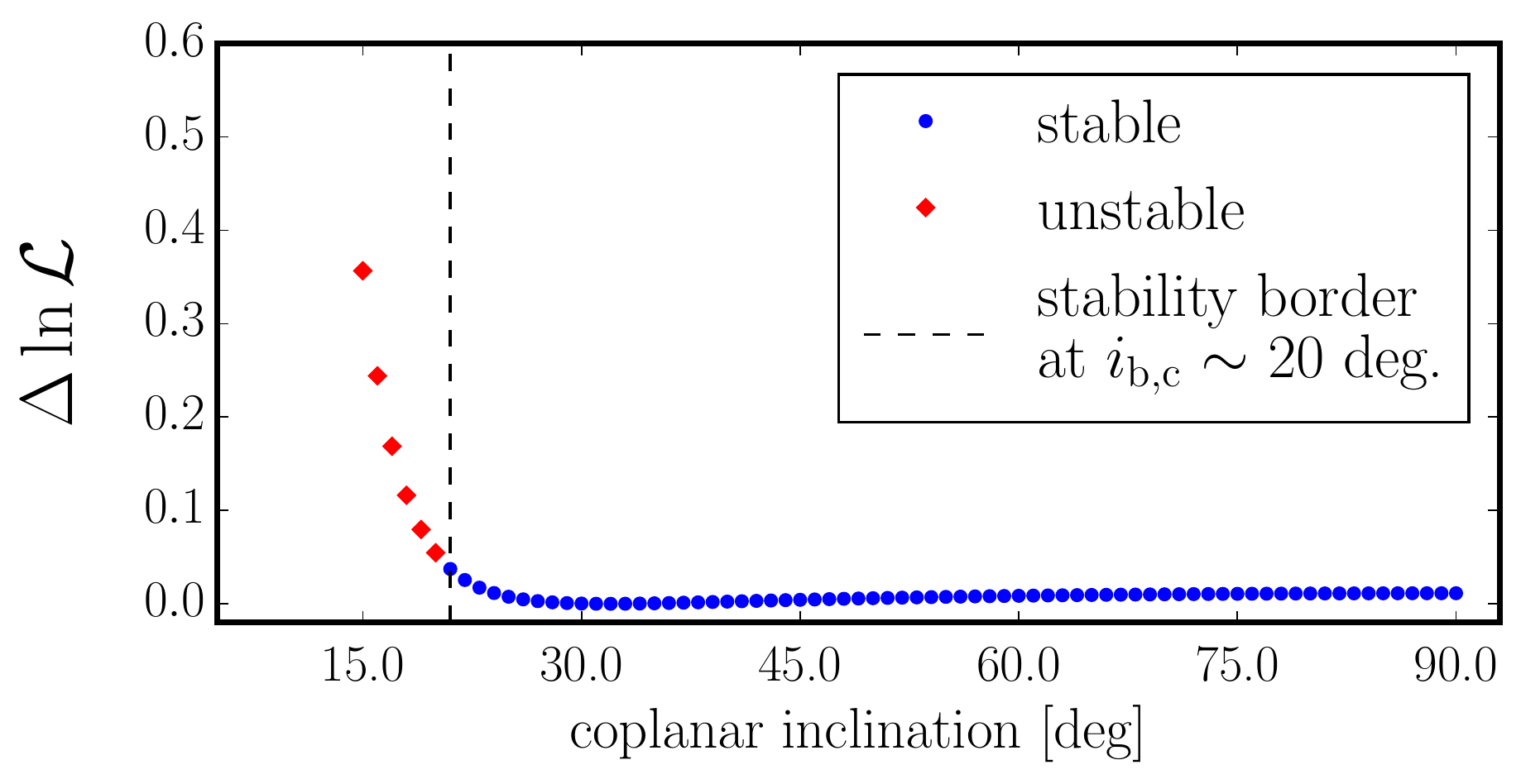}

\end{array} $
\end{center}

\caption{ Coplanar dynamical fits with nearly circular orbits as a function of inclination. 
The stability boundary is at $i$ = 20$^\circ$, 
which sets a planetary mass limit of 
approximately three times the minimum planetary mass of 
the edge-on coplanar case (e.g. see Table~\ref{table:mod_stable}).
}  
\label{incl_stab} 
\end{figure}

\begin{figure*}[]
\begin{center}$
\begin{array}{ccc}

\includegraphics[width=9cm]{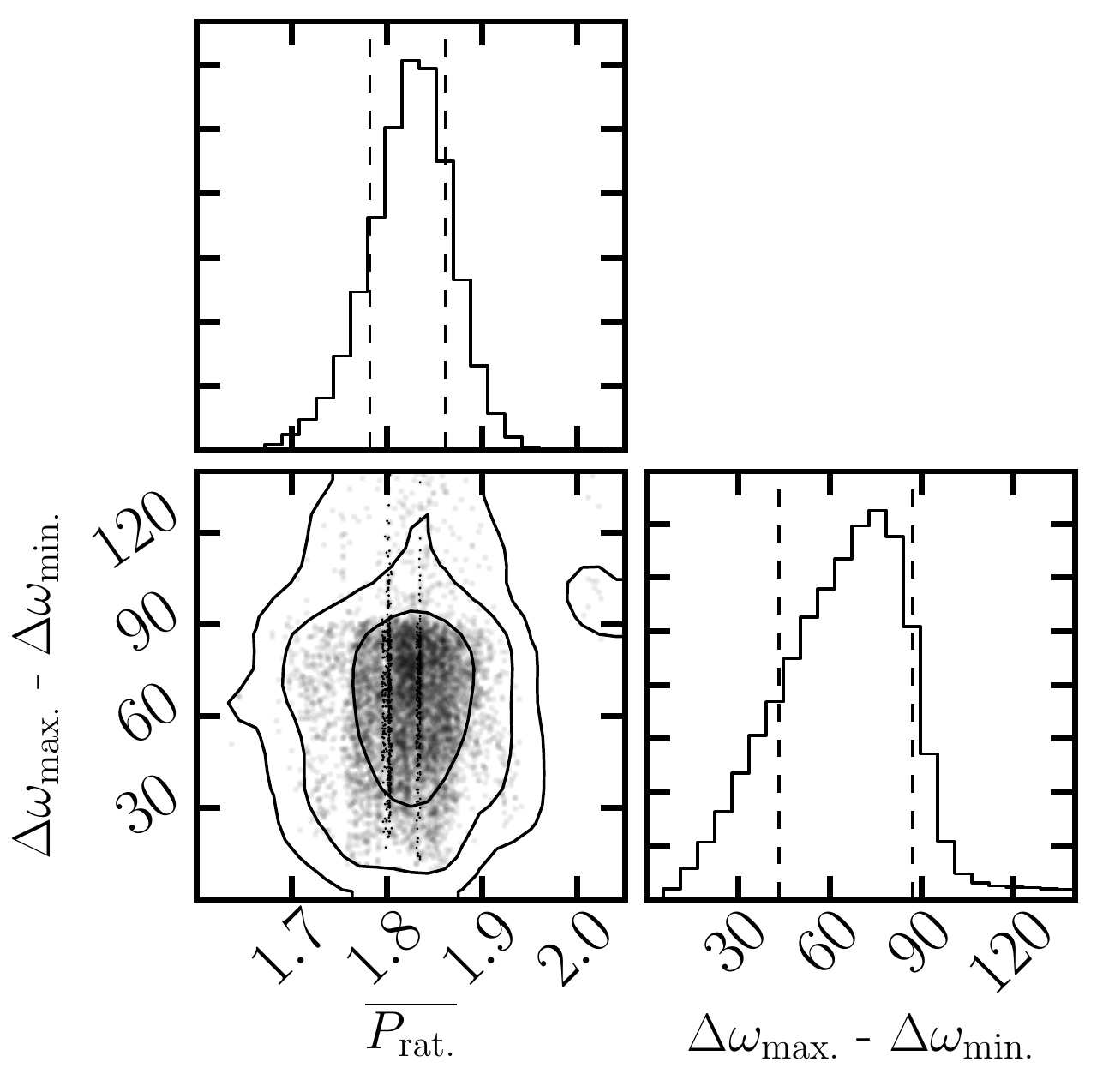}
\put(-80,150){\normalsize $\Delta\omega \sim 0^\circ$}
\includegraphics[width=9cm]{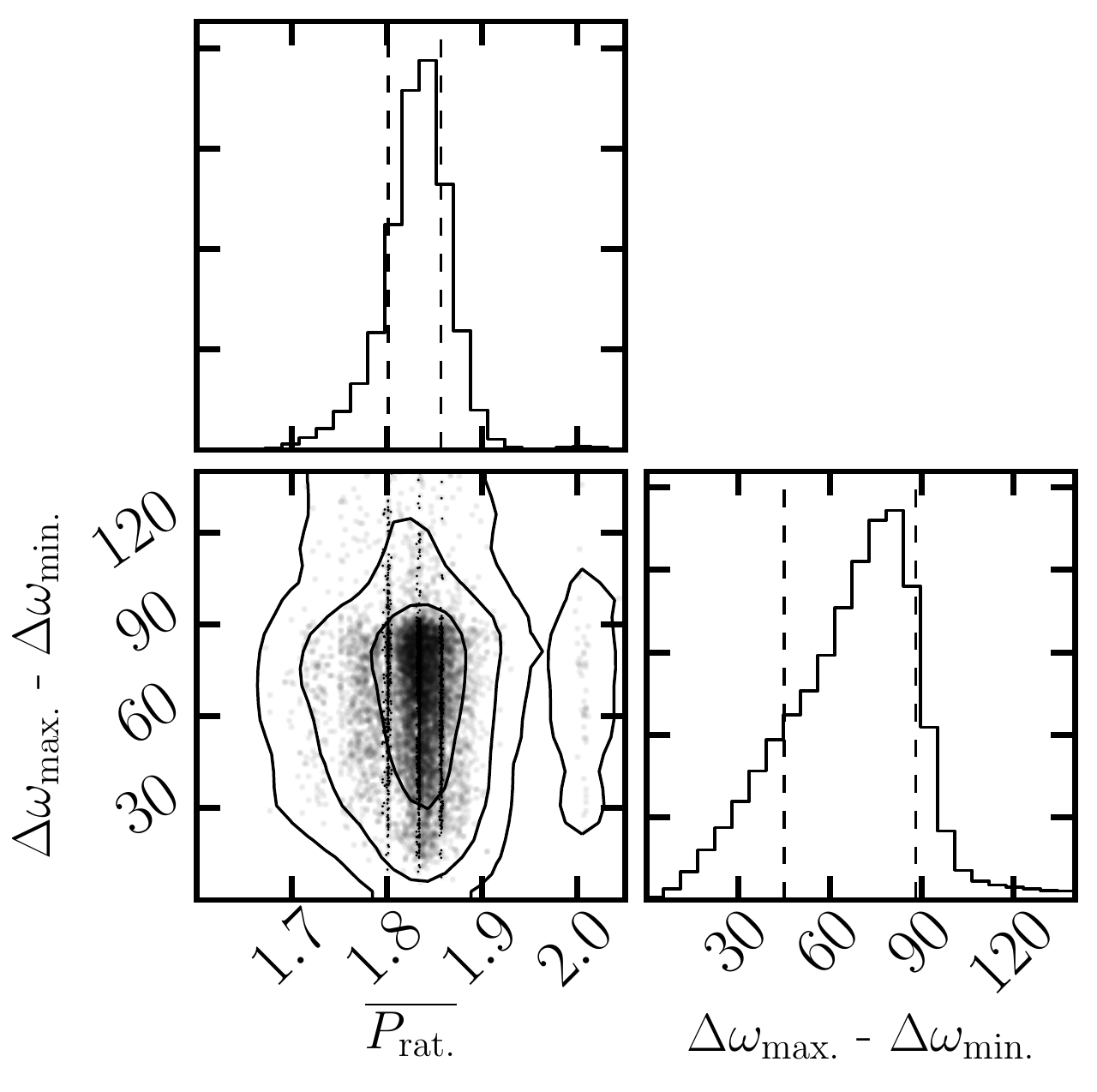}
\put(-80,150){\normalsize $\Delta\omega \sim 180^\circ$}

\end{array} $
\end{center}

\caption{Distribution of the mean period ratio $\overline{P_{\rm rat.}}$ and the amplitude of the apsidal argument $\Delta\omega$
for all dynamical fits that are stable up to 1 Myr (see Fig.~\ref{mcmc}) and consistent with libration of $\Delta\omega$. 
The left panel shows those stable samples whose dynamical properties are consistent with aligned apsidal libration ($\Delta\omega \sim 0^\circ$), while the right panel shows the samples with anti-aligned apsidal libration ($\Delta\omega \sim 180^\circ$).
Both geometries are consistent with  $\overline{P_{\rm rat.}} \sim$ 1.83 (near 11:6 MMR) and 
have similar $\Delta\omega$ libration amplitudes. 
Sample concentrations near the 9:5 and 13:7 MMRs (slightly enhanced) are also within the 1\,$\sigma$ confidence 
levels of the stable posterior distribution.
A small and insignificant fraction of the samples are found in the 2:1 MMR.
}   
\label{res_angle_lib_dist} 
\end{figure*}

Both the Keplerian and dynamical fits reveal a moderate best-fit value
of $e_{\rm c}$ for the eccentricity of the outer planet but with a large uncertainty toward nearly circular orbits. 
This indicates that $e_{\rm c}$ is likely poorly constrained, and 
perhaps this moderate eccentricity value is a result of overfitting.
Indeed, a two-planet Keplerian model with circular orbits (i.e.\ $e_{\rm b,c}$, $\omega_{\rm b,c}$ = 0)  has
$-\ln\mathcal{L}$ = 150.286, meaning that the difference between the  circular and nonciruclar Keplerian
models is only $\Delta\ln\mathcal{L}$ $\approx$ 1.5, which is 
insignificant\footnote{Assuming that the 2$\Delta\ln\mathcal{L}$ follows a 
$\chi^2$ distribution for nested models with $k$ = 4 degrees of freedom \citep[e.g.\ see][]{Baluev2009}.
Alternatively, the relative probability $R$ = $e^{\Delta\ln\mathcal{L}}$ between two competing models
constructed from the same parameters (relevant for the MCMC test) requires $\Delta\ln\mathcal{L} > 5$ to claim significance \citep[see][]{Anglada2016}
}.
Thus, we conclude that neither the Keplerian nor the dynamical model
applied to the current RV data can place tight constraints on the orbital eccentricities. 

The inclusion of the planetary inclinations $i_{\rm b,c}$ and $\Delta\Omega$ 
(i.e.\ $\Delta i$) in the dynamical modeling was also not justified. 
We find that all dynamical fits with coplanar inclinations in the range of 5$^\circ$ $<$  $i_{\rm b,c}$ $<$ 90$^\circ$ 
are practically indistinguishable from the coplanar and edge-on configuration 
($i_{\rm b,c}$ = 90$^\circ$ and $\Delta\Omega$ = 0$^\circ$).
Fits with $i_{\rm b,c}$ $<$ 5$^\circ$ are of poorer quality due to the increased 
dynamical masses of the companions and the strong gravitational interactions, 
which are not consistent with the data anymore. 
Therefore, any further attempt to constrain the 
planetary inclinations as coplanar or mutually inclined leads to inconclusive results.


Overall, our orbital parameter estimates show that taking the planetary gravitational interactions  
into account in the fitting does not yield a significant improvement over the Keplerian model. 
Yet we decide to rely on the dynamical modeling, 
since this scheme provides physical constraints on the 
derived orbital parameters and naturally penalizes highly 
unstable (unrealistic) orbital solutions.


\section{Dynamical characterization and stability}
\label{dynamical_stability}

\subsection{Numerical setup}
\label{Nsetup} 
The long-term stability and dynamical properties of the HD\,202696 system are tested using the SyMBA $N$-body symplectic integrator \citep{Duncan1998},
which was modified to work in Jacobi coordinates \citep[e.g.,][]{LeeM2003}. 
We integrate each MCMC sample for a maximum of 1 Myr with a time step of 4~days. 
In case of planet-planet close encounters, however, 
 SyMBA automatically reduces the time step to
ensure an accurate simulation with high orbital resolution.
 SyMBA also checks for planet-planet or planet-star collisions 
or planetary ejections and interrupts the integration if they occur.
A planet is considered lost and the system unstable if, at any time:
(i) the mutual planet-planet separation is below the sum of their physical radii (assuming Jupiter mean density),
i.e.\ planets undergo collision;
(ii) the star-planet separation exceeds two times the initial semi-major axis of the 
outermost planet ($r_{\rm max} > 2\times a_{\rm c~in.}$), which we define as planetary ejection; or 
(iii) the star-planet separation is below the physical stellar radius ($R \approx$ 0.03 au), 
which we consider as a collision with the star.
All these events are associated with large planetary eccentricities 
leading to crossing orbits, close planetary encounters, rapid exchange of energy and angular momentum, 
and eventually instability.
Therefore, these somewhat arbitrary stability criteria are efficient to detect unstable configurations and save CPU time.

\subsection{Stable orbital space}
\label{Stab_mcmc} 
 
Figure~\ref{mcmc} shows the posterior MCMC distribution of the fitted parameters with a
dynamical modeling scheme whose orbital configuration is edge-on, coplanar, and stable for at least 1 Myr. 
The histogram panels on the top in Figure~\ref{mcmc} provide a comparison 
between the probability density distribution 
of the overall MCMC samples (black) and the stable samples (red) for each fitted parameter. 
The two-dimensional parameter distribution panels represent all possible parameter correlations with respect to the 
 best dynamical fit from Table~\ref{table:orb_par_stable}, whose position is marked with blue lines.
The black 2D contours are constructed from the overall MCMC samples (gray)
and indicate the 68.3\%, 95.5\%, and 99.7\% confidence levels (i.e.\ 1$\sigma$, 2$\sigma$ and 3$\sigma$). 
For clarity, in Fig.~\ref{mcmc} the stable samples are overplotted in red.

We find that $\sim$46\% of the MCMC samples are stable. 
The histograms in Figure~\ref{mcmc} clearly show that the distributions of all MCMC samples 
and their stable counterparts are in mutual agreement, 
peaking near the position of the best dynamical fit. 
The main exception is the distribution of $e_{\rm c}$, 
which is bimodal with a stronger peak suggesting nearly circular orbits and a smaller 
peak at somewhat moderate eccentricities from where the best dynamical fit originates.
Here $P_{\rm b}$ and $\omega_{\rm b}$ also show bimodal distributions, 
which are related to the bimodal distribution of $e_{\rm c}$.
As we discussed in Section~\ref{kep_dyn_analysis}, however, the best-fit eccentricities 
are statistically insignificant, and a simpler two-planet model with circular 
orbits is equally good in terms of $-\ln\mathcal{L}$. 
This indicates that the best-fit solution is likely a global $-\ln\mathcal{L}$ 
minimum, which lies in a smaller and isolated orbital parameter space with moderate $e_{\rm c}$.
This minimum and its surroundings, however, are highly unstable and therefore unlikely to represent the HD\,202696 system configuration.
Thus, running an MCMC test reveals a larger, stable, and statistically 
confident orbital parameter space toward $e_{\rm b,c}$ $\rightarrow$ 0, where
the stable fraction of the sample distribution is $\sim$75\%.
From the long-term dynamical analysis of the posterior MCMC samples, we 
can place a boundary at planetary eccentricities $e_{\rm b,c}$ $\leq$ 0.1, 
while larger eccentricities are very unlikely.
The overall posterior distribution of $\omega_{\rm b,c}$ and $M_{\rm b,c}$ 
are consistent with the best-fit estimate, but the stability 
test cannot constrain these parameters. Because of the low eccentricities of the 
stable solutions, these parameters are ambiguous and  
can be found anywhere between 0$^\circ$ and 360$^\circ$.
However, the orbital mean longitudes 
$\lambda_{\rm b,c}$ = $\omega_{\rm b,c}$ + $M_{\rm b,c}$ 
are rather well constrained to $\sim$ $\pm$ 50$^\circ$ (see Table~\ref{table:orb_par_stable}).

Given the stability constraints from our MCMC test, the most realistic 
configuration of the HD\,202696 system is represented by the parameter distribution of the stable fits.  
In Table~\ref{table:mod_stable} we present our final orbital estimates of the two-planet system assuming coplanar and edge-on 
($i_{\rm b,c}$ = 90$^\circ$, $\Delta\Omega$ = 0$^\circ$) configurations.
These parameters are adopted from the maximum value of the stable parameter probability density function, 
while the uncertainties are  the 0.16 and 0.84 quantiles of the distribution. 
Our new orbital estimate is consistent with circular orbits 
$e_{\rm b} = 0.011_{-0.011}^{+0.078}$ and $e_{\rm c} = 0.029_{-0.012}^{+0.065}$ but otherwise consistent with the 
best two-planet dynamical fit shown in Table~\ref{table:orb_par_stable}.

Integrating this configuration for 100 Myr reveals a stable 
planetary system with a mean period ratio $\overline{P_{\rm rat.}}$ $\approx$ 1.83
and orbital eccentricities oscillating 
in opposite phase in the range from 0 to 0.05 for $e_{\rm b}$ 
and 0 to 0.04 for $e_{\rm c}$.
This configuration librates mostly in an anti-aligned geometry 
with a secular apsidal angle  
$\Delta\omega = \omega_{\rm b}-\omega_{\rm c}$ $\sim$ 180$^\circ$ 
exhibiting a large libration amplitude of $\pm$ 80$^\circ$. 

\begin{table}[ht]

\centering   
\caption{Mode of the dynamically stable MCMC distribution for the HD~202696 system 
(peak of the red histograms from Fig.~\ref{mcmc}) }   
\label{table:mod_stable}      

\begin{tabular}{lrrr}     

\hline\hline  \noalign{\vskip 0.7mm}

\multicolumn{4}{c}{ Adopted Two-planet Coplanar Edge-on  }           \\
\multicolumn{4}{c}{ Configuration Based upon Stability}           \\

\hline \noalign{\vskip 0.7mm}  

Parameter &\hspace{0.0 mm} & HD~202696 b & HD~202696 c  \\
\hline\noalign{\vskip 0.5mm}

Semi-amplitude $K$  [m\,s$^{-1}$]                        &  & 34.1$_{-2.9}^{+2.3}$          &  25.3$_{-3.3}^{+2.1}$        \\
Period $P$ [days]   		        	          &  & 517.8$_{-3.9}^{+8.9}$        &  946.6$_{-20.9}^{+20.7}$      \\  
Eccentricity $e$                                       &  & ~~0.011$_{-0.011}^{+0.078}$   &  ~~0.028$_{-0.012}^{+0.065}$ \\
Arg. of periastron $\omega$ [deg]                            &  & ~~201.6$_{-98.5}^{+97.8}$    &  ~~136.8$_{-62.7}^{+143.0}$       \\   
Mean anomaly $M_0$ [deg]                               & &  ~~79.2$_{-28.1}^{+198.6}$    &  180.0$_{-108.9}^{+98.9}$    \\  \noalign{\vskip 0.9mm} 
Mean longitude $\lambda$ [deg]                           & &  ~~298.8$_{-14.6}^{+70.0}$    &  262.8$_{-53.3}^{+37.8}$    \\ 
Semi-major axis $a$ [au]                                  &  & 1.566$_{-0.007}^{+0.016}$     &  2.342$_{-0.035}^{+0.034}$                     \\  
Dyn. mass $m$ [$M_{\mathrm{Jup}}$]                  & &  1.996$_{-0.100}^{+0.220}$     &  1.864$_{-0.227}^{+0.177}$                    \\ \noalign{\vskip 0.9mm}

Inclination $i$ [deg]                                 & & 90.0 (fixed)       & 90.0 (fixed)              \\ 
Node $\Delta \Omega$ [deg]                     & &  \multicolumn{2}{c}{0.0 (fixed) }                    \\

RV offset~[m\,s$^{-1}$]        &  & \multicolumn{2}{c}{--14.42$_{-2.22}^{+1.70}$}  \\ \noalign{\vskip 0.9mm} 
RV jitter [m\,s$^{-1}$]                      &  & \multicolumn{2}{c}{9.36$_{-0.57}^{+2.03}$}   \\
\noalign{\vskip 0.5mm}

\hline\hline\noalign{\vskip 1.2mm}  

%
%
%

\end{tabular}  

\tablecomments{\small Same comments as in Table~\ref{table:orb_par_stable} }

\end{table}

The long-term stability of the orbital estimate shown in Table~\ref{table:mod_stable} boosts the confidence that the most likely orbital configuration of the HD\,202696 system has been identified.

\subsection{Stability constraints on the planetary masses}
\label{pl_incl}

While the dynamical modeling of the HIRES Doppler 
measurements of HD\,202696 was unable to qualitatively 
constrain the masses of HD\,202696 b and c, 
we also use the long-term stability of 
fits with forced inclination as an efficient way to limit the range of the possible planetary masses further.
Figure~\ref{incl_stab} shows the results from this test. 
We inspect the range of inclinations from $i$ = 90$^\circ$ to 5$^\circ$. 
To ensure enhanced stability, we adopt the planetary eccentricity estimates from 
Table~\ref{table:mod_stable} which for each adopted $i$ remained fixed in the fit. 
The small $\Delta\ln\mathcal{L}$ of the dynamical models in the range of $i$ = 90$^\circ$--15$^\circ$ 
shows that these models are statistically equivalent, but the fit quality 
rapidly deteriorates for $i$ $<$ 15$^\circ$ (not shown in Fig.~\ref{incl_stab}). 
This is likely due to the significantly increased planetary masses, 
which may alter the system stability even at lower eccentricities.

Inclination $i$ = 20$^\circ$ seems to be a stability 
boundary of the coplanar case, leaving the companion masses in the planetary regime with a maximum of
about three times the minimum planetary masses of the coplanar edge-on case. 
However, since orbital inclinations near $i=90^\circ$
are statistically more likely, we limit our dynamical analysis of the HD\,202696 system to the coplanar and edge-on configuration.

\subsection{Dynamical properties of the stable MCMC samples  }
\label{dynamics_18}  

\begin{figure*}[btp]
\begin{center}$
\begin{array}{cc}

\includegraphics[width=6cm]{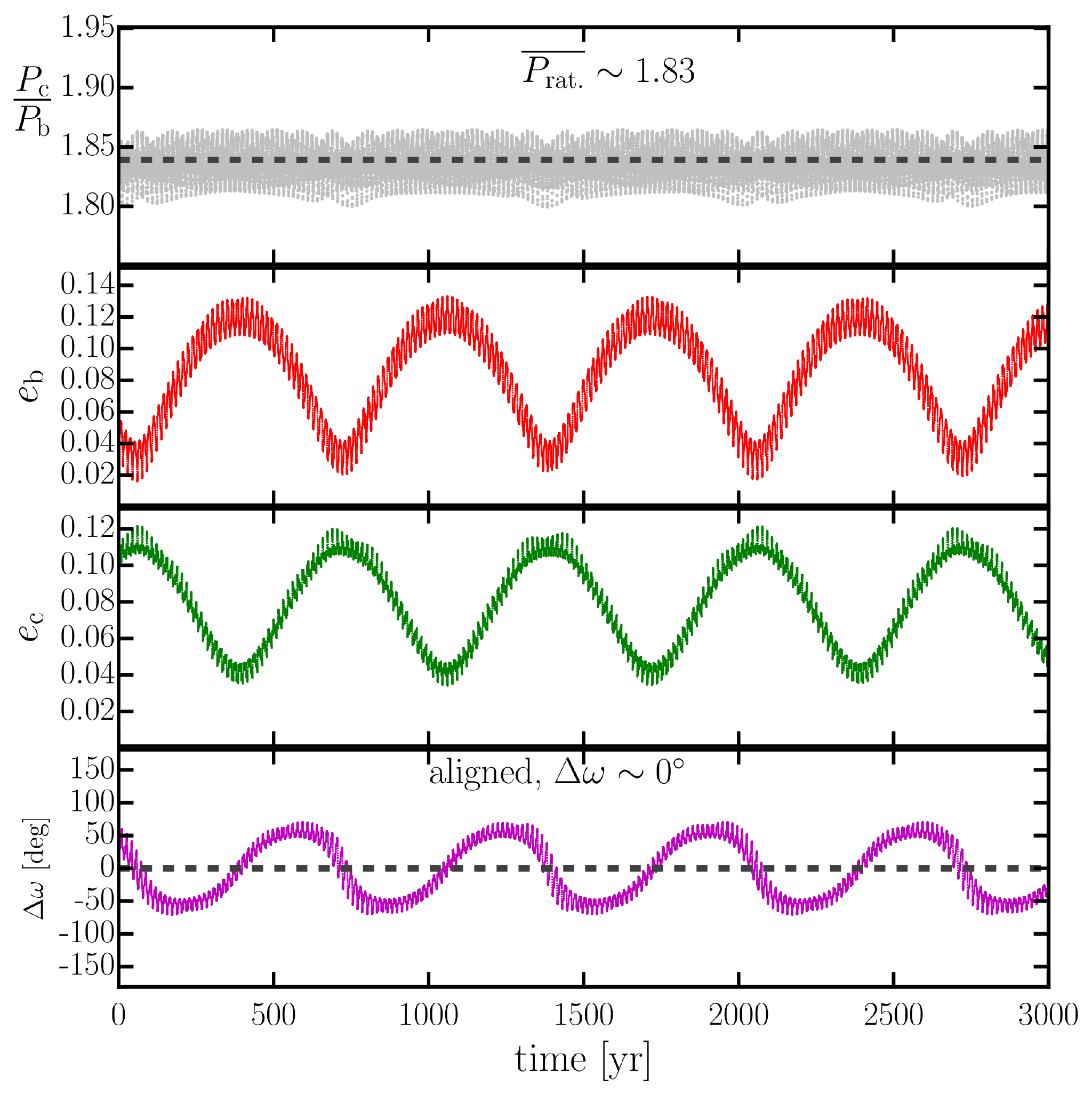}
\includegraphics[width=6cm]{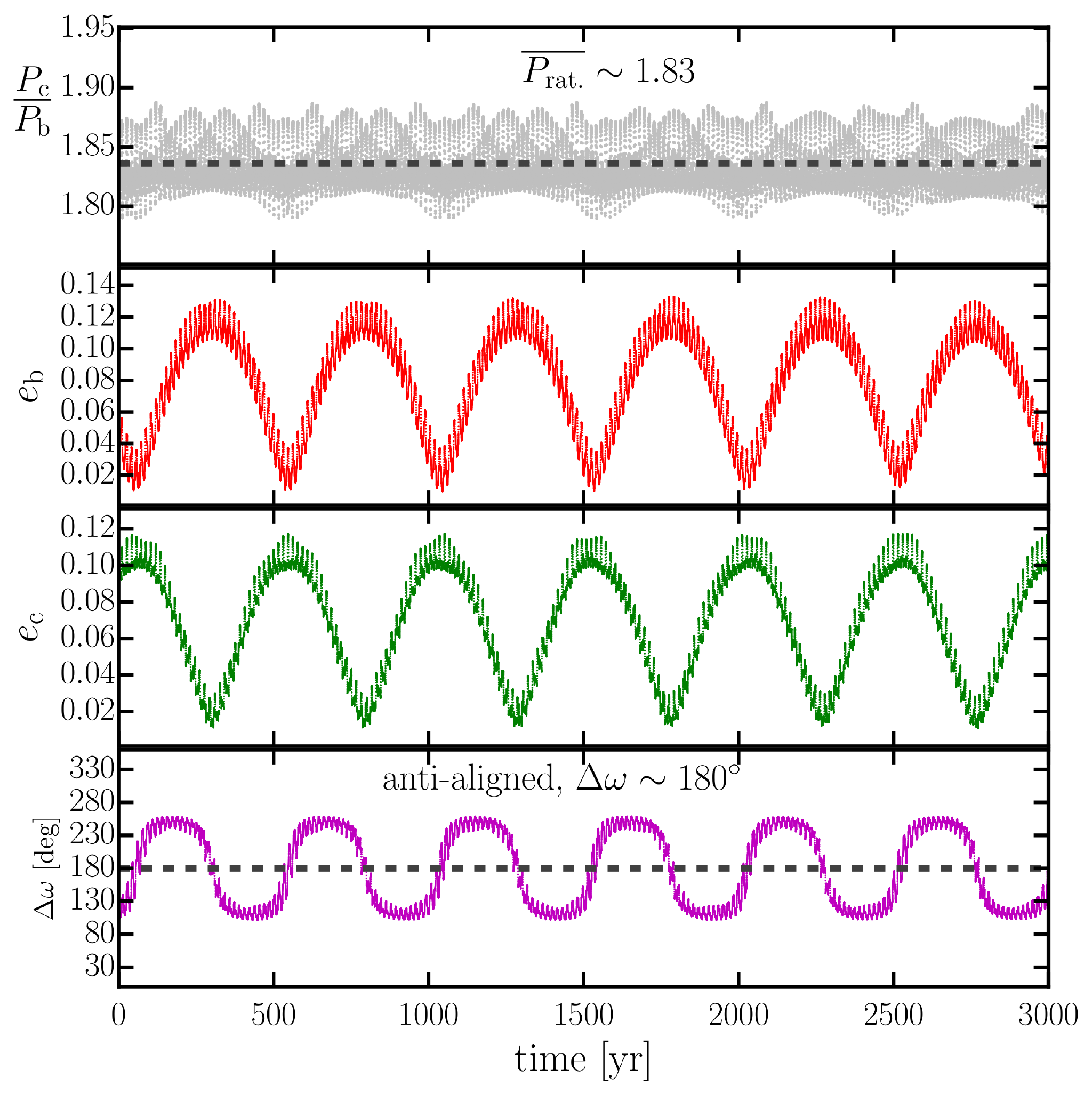}
\includegraphics[width=6cm]{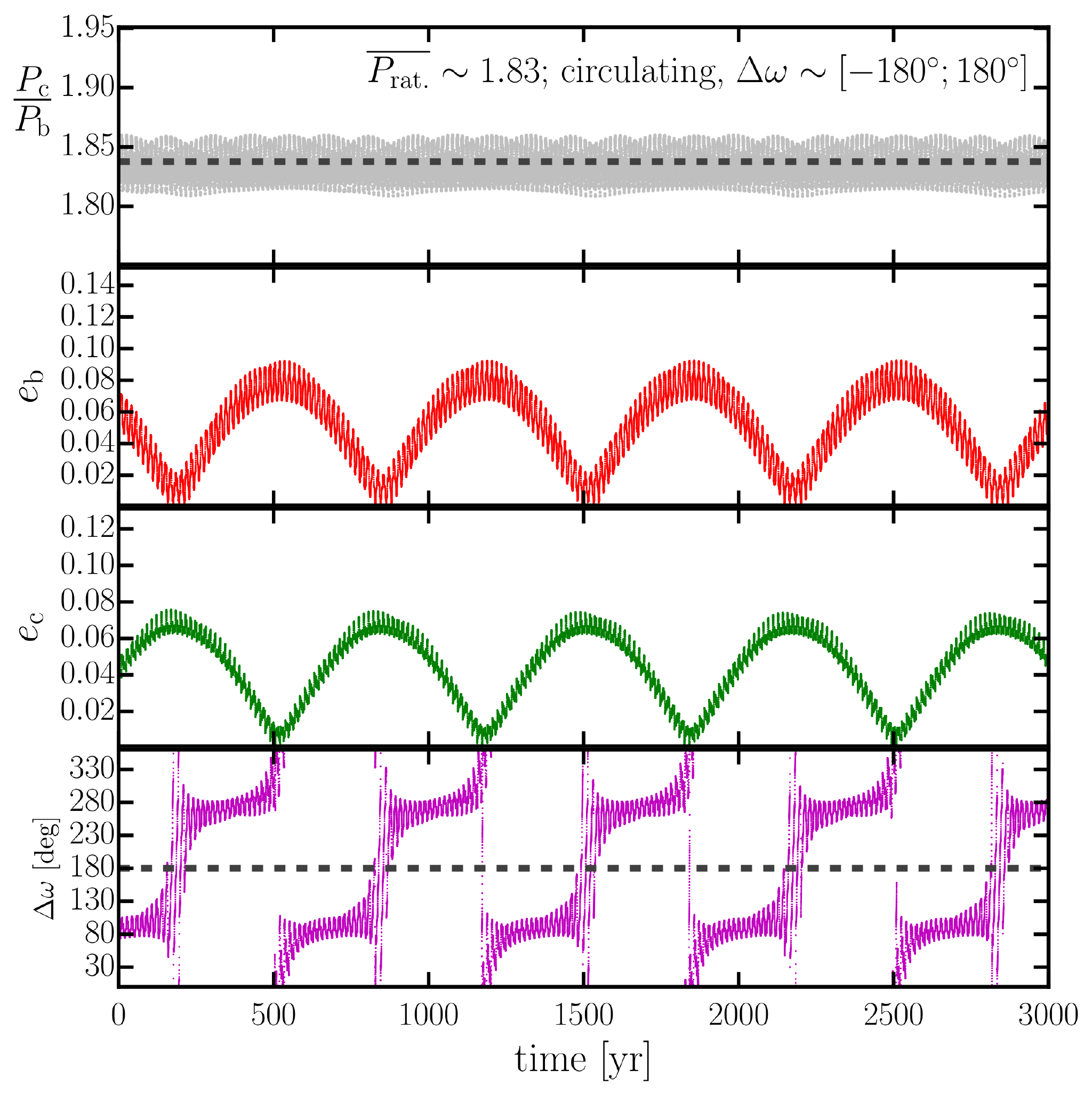}\\

\end{array} $
\end{center}

\caption{Reconstruction of three possible outcomes of the orbital evolution of 
the HD\,202696 system with $\overline{P_{\rm rat.}} \sim$ 1.83, randomly chosen from the stable posterior 
distribution.
Left panel: orbital evolution with $\Delta\omega$ librating around 0$^\circ$.
Middle panel: a case where $\Delta\omega$ librates around 180$^\circ$.
Right panel: a case where $\Delta\omega$  circulates between 0$^\circ$ and 360$^\circ$. 
Note the large amplitude of $P_{\rm rat.}$ in all three cases.
}   

\label{evol_prat1_8} 
\end{figure*}

Overall dynamical analysis of the stable samples reveals that the mean 
period ratio over 1\,Myr of integrations is $\overline{P_{\rm rat.}}$ = 1.83$_{-0.05}^{+0.04}$,
which is consistent with the period ratio from which the system is 
initially integrated, ${P_{\rm rat. in.}}$ = 1.82$_{-0.03}^{+0.03}$. 
This implies that the stable MCMC orbital configurations remain
regular and well separated at any given time of the stability test.
From these stable samples, $\sim$29\% are found with an aligned apsidal libration 
($\Delta\omega \sim 0^\circ$), $\sim$34\% in 
an anti-aligned apsidal libration ($\Delta\omega \sim 180^\circ$; similar to the orbital evolution of the 
long-term stable configuration presented in Table~\ref{table:mod_stable}), 
and  $\sim$37\%  in apsidal circulation ($\Delta\omega$ in the range $0^\circ$--$360^\circ$),
respectively. In fact, the latter case usually involves $e_b$ decreasing from near its maximum value 
to near zero and $e_c$ increasing from near zero to near its maximum value at $\Delta\omega \approx 90^\circ$,
and vice versa at $\Delta\omega \approx 270^\circ$, with episodes of rapid circulation when $e_{\rm b}$ or $e_{\rm c}$
is nearly zero (see right panel of 
Fig.~\ref{evol_prat1_8}). This type of evolution is consistent with a phase space dominated 
by large libration islands about $\Delta\omega = 0^\circ$ and $180^\circ$ and having only a narrow region of circulation (see, e.g., \citealt{LeeM2003}).


Figure~\ref{res_angle_lib_dist} shows the distribution of the mean period ratio $\overline{P_{\rm rat.}}$ and the libration amplitudes
of $\Delta\omega$ for the stable dynamical samples with aligned apsidal libration $\Delta\omega \sim 0^\circ$ (left panel) and 
with anti-aligned apsidal libration $\Delta\omega \sim 180^\circ$ (right panel).
From Fig.~\ref{res_angle_lib_dist} it is clear that both alignment and anti-alignment have similar libration amplitudes of $\Delta\omega$ peaking around 
70$^\circ$ and with a sharp border at 90$^\circ$, beyond which the system is involved in apsidal circulation. 
An interesting pileup in the posterior distribution of
$\overline{P_{\rm rat.}}$ can be seen around 1.80, 1.83, and 1.86,
which correspond to the 9:5, 11:6, and 13:7 MMRs.
We inspected these possible high-order MMRs for librating resonance 
angles,\footnote{For the $m$:$n$ MMR, the resonance angles are defined as $\theta_{k=1,m-n} =  n\lambda_{\rm b} + (n-k)\varpi_{\rm b}  - m\lambda_{\rm c} -  
(1-k)\varpi_{\rm c}$, where $\varpi$ and $\lambda$ are the planetary 
longitude of periatron and the mean longitude, respectively. 
At least one resonance angle $\theta_{k}$ must librate to claim a MMR configuration.} but we did not detect any.
Despite the apsidal libration of the $\Delta\omega$ in $\sim$60\% of the 
configurations, which is suggestive of resonant behavior,  
the system seems to be out of any MMR. 
In the $\Delta\omega \sim 180^\circ$  case, a small number of 
system configurations are consistent with a 2:1 period ratio, but these samples have 
statistically poorer $-\ln\mathcal{L}$ and are thus unlikely to represent the HD\,202696 system.

Given the large fraction of nonresonant stable configurations, we 
concluded that the system HD\,202696 is most likely dominated by 
secular interactions at low eccentricities.
Figure~\ref{evol_prat1_8} shows the orbital evolution of three confident 
stable configurations that have a mean period ratio of 
$\overline{P_{\rm rat.}} \sim$ 1.83,
exhibiting $\Delta\omega \sim 0^\circ$ and $\Delta\omega \sim 180^\circ$, 
and $\Delta\omega$ circulating, respectively.  
The various stable configurations show libration amplitudes 
in their eccentricities between 0 and 0.15 but 
with mean values consistent with the posterior  
eccentricity distribution. The secular time scales of these 
oscillations strongly depend on the initial orbital geometries
and planetary masses, ranging mostly between $\sim$200 and $\sim$1000 yr.

\section{Discussion}
\label{Discussion}

\subsection{Tight Jovian Pairs around Evolved Stars}
\label{discusion1}  

The HD\,202696 system is another example of a growing number of 
multiplanetary systems around evolved intermediate-mass stars 
with semi-major axes close to the critical stability boundary.
Including HD\,202696, there are
now seven tightly packed 
planetary systems with period ratios smaller than or about 2:1.

Earlier examples are the 24 Sex and HD 200964 systems \citep{Johnson2011}.
A 1.5$M_\odot$ subgiant 24 Sex is orbited by two Jovian planets 
with minimum dynamical masses of $m_{\rm b}$ $\sim$ 2, and
$m_{\rm c}$ $\sim$ 0.9$M_{\rm Jup}$, and periods of 
$P_{\rm b}$ $\approx$ 450 and $P_{\rm c}$ $\approx$ 883\,days.
The best-fit period ratio for the 24 Sex system is slightly below 2:1, 
but the planets are almost certainly involved in a 2:1 MMR \citep{Johnson2011,Wittenmyer2012}.
The HD\,200964 system is similar to 
24~Sex in terms of stellar mass and minimum planetary masses,
but it has a much more compact orbital configuration, with periods of 
$P_{\rm b}$ $\approx$ 610 and $P_{\rm c}$ $\approx$ 830\,days. 
The HD\,200964 system is only stable if the orbits are in a 4:3 MMR \citep{Wittenmyer2012,Santos2015}.

The 1.5$M_\odot$ subgiant star HD\,5319 is another 4:3 MMR candidate host \citep{Robinson2007, Giguere2015}. 
With minimum planetary masses of 
$m_{\rm b} \sin i$ $\sim$ 1.8$M_{\rm Jup}$,
$m_{\rm c} \sin i$ $\sim$ 1.2$M_{\rm Jup}$, and periods of 
$P_{\rm b}$ $\approx$ 640\,d and $P_{\rm c}$ $\approx$ 890\,d, respectively,
this system is dynamically very challenging. 
\citet{Giguere2015} showed that, although most of the possible configurations of HD\,5319 are unstable, 
stable solutions at the exact 4:3 MMR do exist. Therefore, the 
4:3 MMR seems like the most reasonable explanation of the coherent Doppler signal of HD\,5319.

A 1.7$M_\odot$ slightly evolved star, HD\,33844 was reported by \citet{Wittenmyer2016}
to host two Jovian-mass planets with minimum dynamical masses $m_{\rm b}$ $\sim$ 1.9, and
$m_{\rm c}$ $\sim$ 1.7$M_{\rm Jup}$, 
and periods of $P_{\rm b}$ $\approx$ 550, $P_{\rm c}$ $\approx$ 920\,days, respectively.
These authors concluded that the planetary pair is most likely trapped in the 5:3 MMR, 
but stable solutions consistent with 8:5 and 12:7 MMR were also 
found within the 1$\sigma$ uncertainty in the semi-major axes.

A 1.8$M_\odot$ giant, HD\,47366 has two nearly 
equal-mass planetary companions \citep{Sato2016}. 
With $m_{\rm b} \sin i$ $\sim$ 1.8, $m_{\rm c} \sin i$ $\sim$ 1.9$M_{\rm Jup}$ 
and periods of $P_{\rm b}$ $\approx$ 363 and $P_{\rm c}$ $\approx$ 685\,days, the system is dynamically challenging.
Within the uncertainties of the orbital parameters given, the initial period ratio is close to 15:8, 
where the system could be stable if the planets are on circular orbits. 
\citet{Sato2016} also proposed that HD\,47366 could be locked in a retrograde 2:1 MMR. 
Tightly packed retrograde configurations are, however, difficult to reconcile with current planet formation scenarios. 
\citet{Marshall2018} showed that alternatives to these configurations for 
HD\,47366 exist, including stable prograde solutions in the 2:1 MMR.

\begin{figure}[btp]
\begin{center}$
\begin{array}{cc} 

\includegraphics[width=9cm]{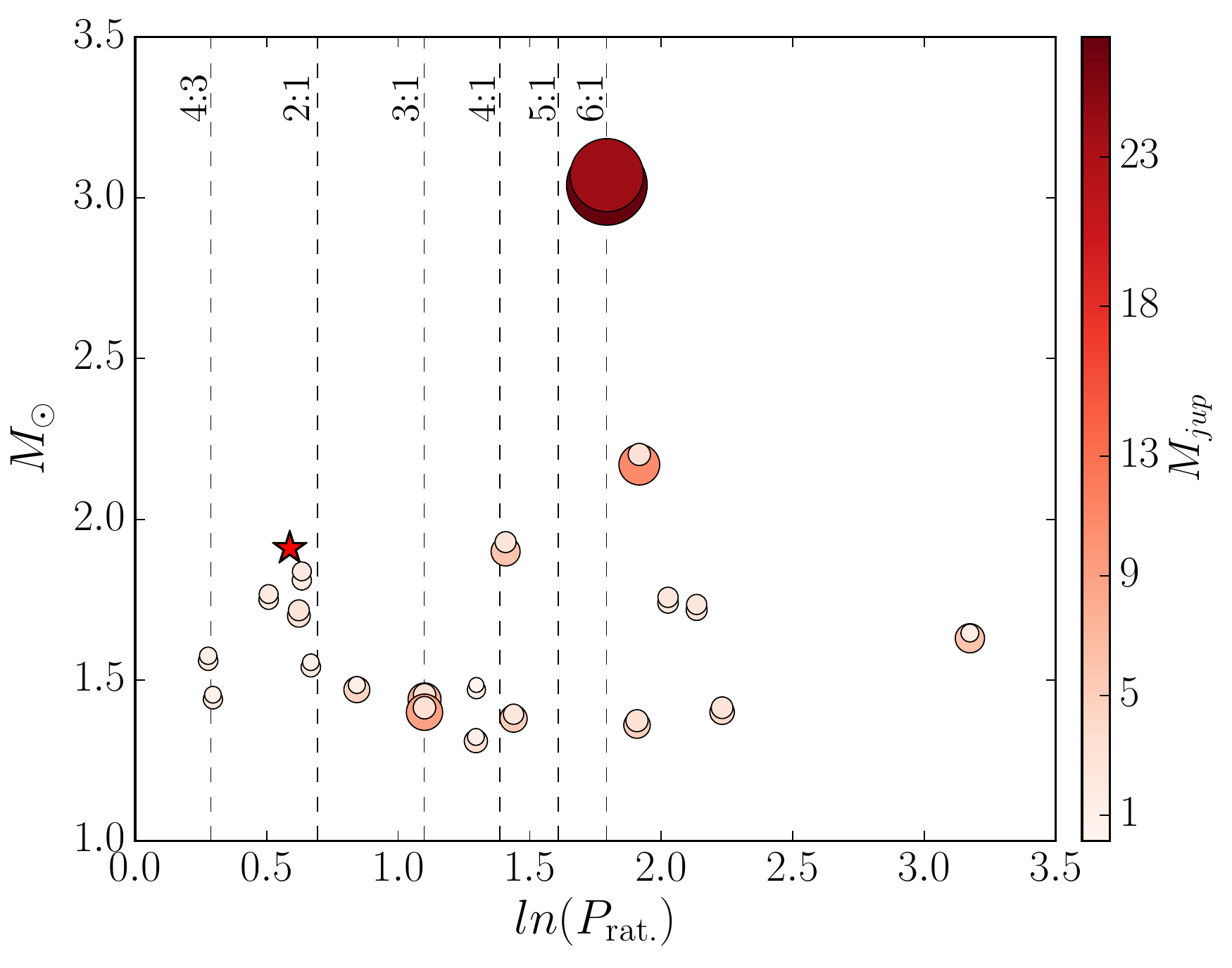}

\end{array} $
\end{center}

\caption{Period ratio of known multiplanetary systems discovered via the Doppler method and
orbiting stars with estimated masses larger than 1.3\,$M_{\odot}$.
Each planetary companion is plotted at the published best-fit period 
ratio for the system, with a symbol that is color-coded and scaled in size with its reported mass.
Six dynamically challenging systems are known to orbit at or below the 2:1 MMR.
A new member of this sample with a period ratio of 11:6, HD\,202696 is indicated in the plots with a red star symbol. See text for details.
}

\label{prat_stmass} 
\end{figure}

The $\eta$ Ceti system is composed of a 
giant star orbited by a planet pair with minimum dynamical masses of
$m_{\rm b}$ $\sim$ 2, $m_{\rm c}$ $\sim$ 3$M_{\rm Jup}$, and nominal best-fit periods of $P_{\rm b}$ $\approx$ 400 and $P_{\rm c}$ $\approx$ 750\,days, respectively.
\citet{Trifonov2014} found two possible stability regimes 
for $\eta$ Ceti: either it is a strongly interacting two-planet system locked in an anti-aligned 
2:1 MMR, or, with lower probability, it could be a low-eccentricity two-planet system dominated by secular 
interactions close to a high-order MMR like 9:5, 11:6, 13:7, or 15:8, resembling to some extent the HD\,202696 system.

Figure~\ref{prat_stmass} shows all known RV multiple-planet systems around intermediate-mass stars 
with masses larger than 1.3$M_\odot$\footnote{Taken from \url{http://exoplanet.eu}, regardless 
of their spectral type and evolutionary stage.}. The stellar host mass is shown as a function of the logarithm of the system's period ratio.
A total of 21 planetary\footnote{As can be seen from Fig.~\ref{prat_stmass}
one of the systems is in fact a brown dwarf pair. This is the $\nu$ Oph system locked in a 6:1 MMR
\citep[][Quirrenbach et al.\ A\&A in press.]{Quirrenbach,Sato2012} } 
systems are shown in Fig.~\ref{prat_stmass}. 
The seven planetary systems found between the 2:1 and 4:3 first-order MMRs (and
briefly described above) represent 33\% of all known massive 
multiple-planet systems around intermediate-mass stars. 
This fraction is rather large given the broad range of the observed planetary period ratios. 
The planet pair occurrence rate at low period ratios, of course, 
could be a result of a selection effect, since these systems are likely to be 
discovered first by precise Doppler surveys. 
However, a more intriguing possibility is that the increasing number of these tight 
planetary pairs indicates the existence of a massive multiple-planet system population, 
some of which have successfully broken the 2:1 MMR during the planet migration phase.

 It should be noted that a strongly interacting 2:1 MMR system 
 could, in principle, exhibit prominent oscillations of their semi-major axis (and period ratio). 
 The secular oscillation time scales are usually longer 
 than the temporal baseline of the Doppler measurements and 
 therefore normally only marginally detected in the best-fit modeling. 
 An example is the $\eta$ Ceti system, where best-fit dynamical modeling of the osculating orbits
 suggests $P_{\rm rat.}$ $\sim$ 1.86, but the stable solutions of the system have
 $\overline{P_{\rm rat.}}$ = 2:1, with an MMR libration amplitude of 
 $\Delta\overline{P_{\rm rat.}} \sim 0.2$ and time scales of $\sim$500\,yr \citep{Trifonov2014}. 
 In this context, the 24~Sex, HD\,47366, and even the HD\,202696 system could in
 fact be true 2:1 MMR systems with oscillating period ratios, observed slightly below the 2:1 period ratio at the present epoch.
 More precise Doppler data for these systems taken in the future, with an extended temporal baseline, could
 reveal the true orbital configurations of these systems; we encourage such 
 extended RV campaigns.
 Still, the HD\,33844 system is likely stable at  or near the 5:3 MMR, while the HD\,5319 and the HD\,200964 planetary systems
 can only be stable if they are involved in the 4:3 MMR. The latter systems are dynamically not consistent with the 2:1 MMR  
 and thus indicate that planet pairs could break the 2:1 MMR and settle at other stable orbits 
 in lower-order MMRs.

\subsection{Planetary migration below the 2:1 MMR}
\label{discusion1}

\begin{figure}[btp]
\begin{center}$
\begin{array}{cc} 
\includegraphics[width=9cm]{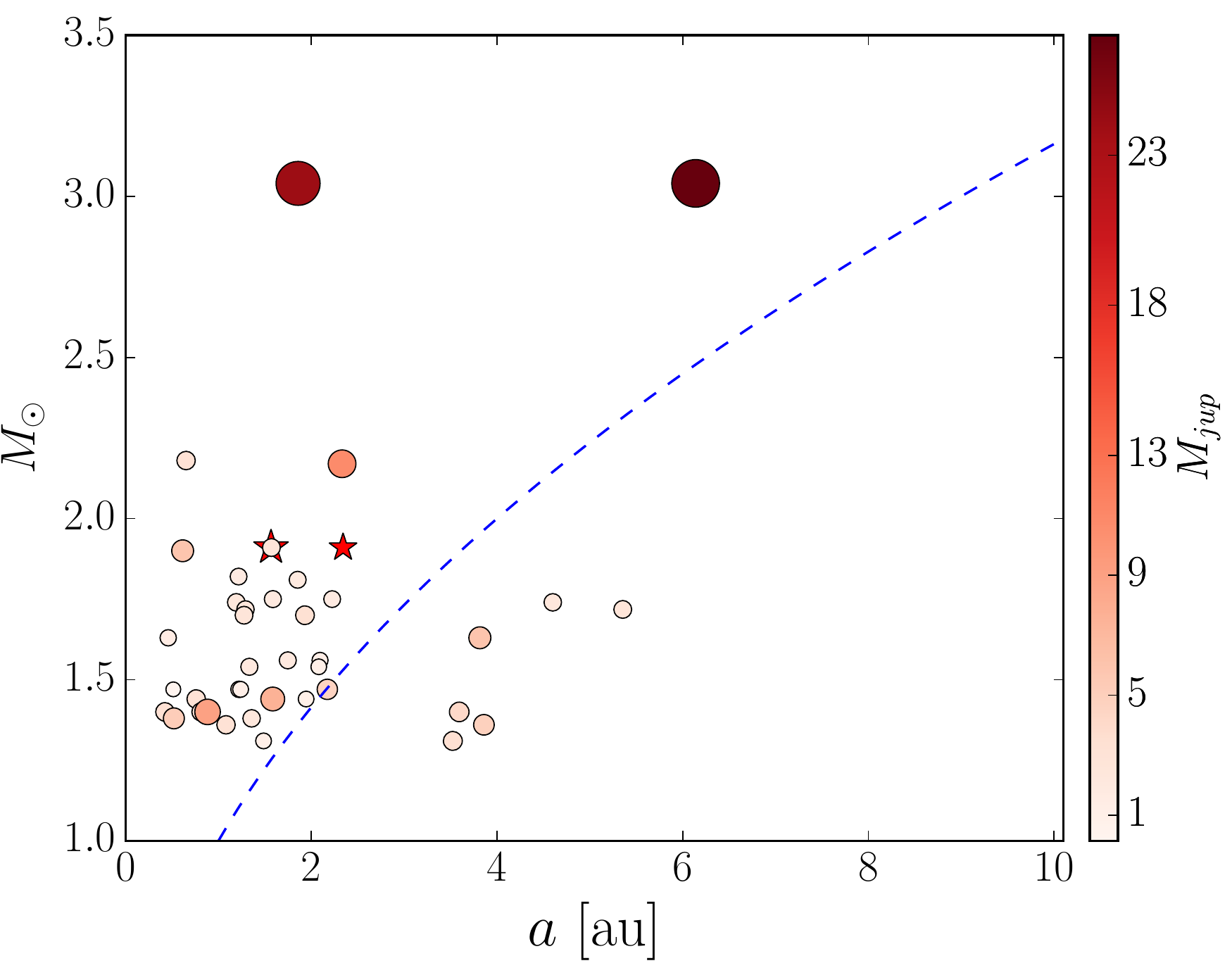}
\end{array} $
\end{center}
\caption{Same as in Figure~\ref{prat_stmass} but as a function of the planetary semi-major axes. 
The dashed blue curve shows the approximate position of the pre-main-sequence "ice line"
for stellar masses between 1 and 3.5\,$M_\odot$. 
Most of the discovered Jovian-mass multiple-planet 
systems around intermediate-mass stars are within the ice line, suggesting that 
these pairs most likely migrated inward together during the disk phase. 
See text for details.  
}\label{a_stmass} 
\end{figure}

The HD\,202696 system, like the majority of the known Jovian-mass planetary systems, is
currently located within the so-called ``ice-line", where it is unlikely to form gas giants. 
This can be easily seen from Fig.~\ref{a_stmass}, which 
shows the planetary systems from Fig.~\ref{prat_stmass} vs.\
their planetary semi-major axes.
The dashed blue line denotes the approximate ice-line radius, 
beyond which it is commonly accepted that planetary embryos are able to
accumulate icy material and grow massive enough to start 
gas accretion from the protoplanetary disk to eventually become Jovian-mass planets.
While accreting disk material, massive planets 
undergo inward migration towards the stellar host before the disk dissipates,
and planets end up at their observed semi-major axes.
We note that the ice line plotted in Fig.~\ref{a_stmass} is calculated  
for the protoplanetary phase 
assuming only stellar irradiation, where the disk 
temperature would be $T_{\rm irr}$ = 150 K \citep{Ida2016}, 
but adopting a stellar mass-luminosity relation of the form 
$L$ = $M^{3.5}$, which is reasonable for main-sequence stars. 
We are aware that this is a crude approximation, but it is sufficient
to demonstrate the generally ``warm orbits" of the discovered Jovian-mass 
planets orbiting intermediate stars.

Fig.~\ref{a_stmass}, in connection with current planet formation theories, suggests that the massive planet pairs have 
formed further out than observed today and have migrated inward together. 
If the migration is convergent, there is a high probability of planets being trapped in an orbital resonance, 
with a final stop at the strong 2:1 MMR \citep[e.g.][]{LeeM2002,Beauge2003}.
Perhaps due to not yet well understood dynamical processes 
between the disk and the planets,
pairs of planets may have broken the 2:1 MMR and migrated further in.

\citet{Andre2016} studied systems similar to 24~Sex and $\eta$ Ceti
 and were able to reproduce the 2:1 MMR in these systems. However, 
 in their simulations, there were also cases in which the planets passed
 through the 2:1 MMR, which made them undergo capture into 
 the destabilizing 5:3 resonance.  \citet{Andre2016} concluded that finding 
 systems in the 5:3 MMR is unlikely. 

In this context, the 4:3 MMR systems are also very challenging from the formation perspective. 
\citet{Rein2012} failed to reproduce the 4:3 MMR of the 
HD\,200964 system using the standard formation scenarios,
such as convergent migration, scattering, and in situ formation.
On the other hand, \citet{Santos2015} were able to closely reproduce the 4:3 resonant dynamics of HD\,200964
by including interactions between type I and type II migration, planetary growth, and stellar evolution.
As pointed out by \citet{Santos2015}, however, 
the 4:3 MMRs capture of the HD\,200964 system occurred only for a very narrow range of initial conditions.
 
In general, massive planets that have reached the gap 
opening mass migrate via type II migration, which
follows the viscous disk evolution \citep{Lin1986}. 
In this case, a two-planet pair would migrate on the 
same time-scale, and it is very hard for the outer 
planet to catch up with the inner planet, let alone to 
break free from a resonance trapping via planet-disk interactions.

However, if the outer planet is less massive and does not open a full gap, 
it can undergo type III migration \citep{Masset2003}. 
This migration mode is even faster than type I migration 
and can easily allow outer planets to catch up with inner, 
more massive planets and be trapped in resonance \citep{Masset2001}. 

This situation has been studied in the framework of the Grand-Tack 
scenario for Jupiter and Saturn \citep{Walsh2011, Pierens2014}. 
In particular, \citet{Pierens2014} found that the inward-migrating 
Saturn could, under certain disk conditions, have broken free from the 2:1 
resonance and migrated closer to Jupiter, where trapping could have occurred, e.g.\ in the 3:2 resonance.

While the planets are still embedded in the gas disk, they will 
continue to accrete gas and grow. 
The outer planet can feed from a larger gas reservoir than 
the inner one because it has direct access to the disk's gas 
supply with the full disk accretion rate. 
The inner planet can only accrete the gas that passed 
the outer planet, which is typically only up to 20\% of 
the disk's accretion rate \citep{Lubow2006}. 
Eventually, the outer planet can then outgrow the inner 
planet if the disk lifetime is long enough.
During their final growth, the dynamical interactions between the giant 
planets change, which could lead to small orbital disturbances breaking 
the final resonance. 

A possible escape from the 2:1 MMR could 
also be a resonance overstability, as in the context of \citet{Goldreich2014}.
In this case, convergent planetary migration with strongly damped 
eccentricities may only lead to a temporal capture at the 2:1 MMR.
However, these scenarios require further investigation, which is 
beyond the scope of this paper.

\section{Summary}
\label{Summary}

The star HD\,202696 is an early K0 giant ($M$ = 1.91$_{-0.14}^{+0.09}$\,$M_\odot$, 
$\log g$ = 3.24 $\pm$ 0.15\,$\mathrm{cm\cdot s}^{-2}$, [Fe/H] = 0.02 $\pm$ 0.04\,dex),
for which HIRES high-precision RV measurements were taken between 2007 and 2014. 
The RV data indicate the presence of at 
least two strong periodicities near 520 and 940\,days,
with no evidence of stellar activity near these periods.
Therefore, the most plausible explanation of the observed RVs  
is stellar reflex motion induced by a pair of Jovian-mass planets, each
with a minimum mass of $\sim$2 $M_{\rm Jup}$.

Dynamical modeling of the RV data is particularly important for the
HD\,202696 system, since planet-planet perturbations 
in this relatively compact system of massive planets cannot be neglected.  
For the orbital analysis, we therefore rely on a self-consistent dynamical 
scheme for the RV modeling, coupled with an MCMC sampling algorithm for parameter analysis. 
Each MCMC orbital configuration was further integrated 
with the SyMBA $N$-body integrator for a maximum of 1 Myr to analyze 
the long-term stability and dynamical properties of the system.

Our long-term stability results yield that approximately 46\%
of the edge-on coplanar ($i$ = 90$^\circ$, $\Delta\Omega$ = 0$^\circ$)
MCMC samples survived 1 Myr. 
The stable parameter space of confident MCMC configurations 
suggests planets on nearly circular orbits with 
(minimum) dynamical masses of $m_{\rm b}$ = 2.0$_{-0.10}^{+0.22}$ 
and $m_{\rm c}$ = 1.86$_{-0.23}^{+0.18}$\,$M_{\rm Jup}$.   
The planetary semi-major axes are $a_{\rm b}$ = 1.57$_{-0.01}^{+0.02}$  
and $a_{\rm c}$ = 2.34$_{-0.04}^{+0.03}$ au, 
forming a period ratio of $\sim$ 11:6, which is preserved 
during the numerical orbital evolution. 
Overall, we find that the HD\,202696 system is most likely dominated by secular 
perturbations near the 11:6 mean-motion commensurability, with no direct
evidence of resonance motions. 

In this context, the HD\,202696 system is not unique. 
It belongs to a population of massive multiplanet systems orbiting evolved 
intermediate-mass stars with
orbital period ratios between the first-order MMRs of 2:1 and 4:3.
These systems impose challenges for planet migration theories,
since during the type II migration phase, they 
unavoidably must have passed through the strong 2:1 MMR, which
requires special conditions for each case and thus in general has a low probability.
Understanding the giant-planet migration into high-order 
MMR commensurability after passing the strong 2:1 MMR 
as in the case of the HD\,202696 system requires more sophisticated 
disk-planet migration simulations, which are beyond the scope of this study.

The discovery of more multiplanet systems around G and K giants 
will be important to better understand planet formation and evolution around more massive and evolved stars.

\acknowledgments

This work was supported by the DFG Research Unit FOR~2544, {\it Blue  Planets around Red Stars}, 
project No.~RE~2694/4-1. S.R.\ further acknowledges the support of the DFG Priority Program SPP 1992, {\it Exploring the Diversity of Extrasolar Planets} (RE~2694/5-1). 
M.H.L. was supported in part by Hong Kong RGC grant HKU 17305015.
B.B.\ thanks the European Research Council (ERC Starting grant 757448-PAMDORA) for their financial support.
This research has made use of the SIMBAD database, operated at CDS, Strasbourg, France.
This work has made use of data from the European Space Agency (ESA) mission
{\it Gaia} (\url{https://www.cosmos.esa.int/gaia}), processed by the {\it Gaia}
Data Processing and Analysis Consortium (DPAC;
\url{https://www.cosmos.esa.int/web/gaia/dpac/consortium}). Funding for the DPAC
has been provided by national institutions, in particular the institutions
participating in the {\it Gaia} Multilateral Agreement.
We also wish to extend our special thanks to those of Hawaiian ancestry on
whose sacred mountain of Maunakea we are privileged to be guests. Without their generous hospitality, the Keck
observations presented herein would not have been possible.
We thank the anonymous referee for the excellent comments that helped to improve this paper.

\software{
CERES pipeline \citep{Brahm2017a}, 
ZASPE \citep{Brahm2017b}, 
astroML \citep{VanderPlas2012}, 
Systemic Console \citep{Meschiari2009},
$emcee$ \citep{emcee}
}

\bibliographystyle{apj}

\bibliography{Trifonov_HD202696}

\begin{thebibliography}{}
\expandafter\ifx\csname natexlab\endcsname\relax\def\natexlab#1{#1}\fi

\bibitem[{{Andr{\'e}} \& {Papaloizou}(2016)}]{Andre2016}
{Andr{\'e}}, Q., \& {Papaloizou}, J.~C.~B. 2016, \mnras, 461, 4406

\bibitem[{{Anglada-Escud{\'e}} {et~al.}(2016){Anglada-Escud{\'e}}, {Amado},
  {Barnes}, {Berdi{\~n}as}, {Butler}, {Coleman}, {de La Cueva}, {Dreizler},
  {Endl}, {Giesers}, {Jeffers}, {Jenkins}, {Jones}, {Kiraga}, {K{\"u}rster},
  {L{\'o}pez-Gonz{\'a}lez}, {Marvin}, {Morales}, {Morin}, {Nelson}, {Ortiz},
  {Ofir}, {Paardekooper}, {Reiners}, {Rodr{\'{\i}}guez},
  {Rodr{\'{\i}}guez-L{\'o}pez}, {Sarmiento}, {Strachan}, {Tsapras}, {Tuomi}, \&
  {Zechmeister}}]{Anglada2016}
{Anglada-Escud{\'e}}, G., {Amado}, P.~J., {Barnes}, J., {et~al.} 2016, \nat,
  536, 437

\bibitem[{{Arenou} \& {Luri}(1999)}]{Arenou1999}
{Arenou}, F., \& {Luri}, X. 1999, in Astronomical Society of the Pacific
  Conference Series, Vol. 167, Harmonizing Cosmic Distance Scales in a
  Post-HIPPARCOS Era, ed. D.~{Egret} \& A.~{Heck}, 13--32

\bibitem[{{Bailer-Jones} {et~al.}(2018){Bailer-Jones}, {Rybizki}, {Fouesneau},
  {Mantelet}, \& {Andrae}}]{Bailer_Jones}
{Bailer-Jones}, C.~A.~L., {Rybizki}, J., {Fouesneau}, M., {Mantelet}, G., \&
  {Andrae}, R. 2018, ArXiv e-prints, arXiv:1804.10121

\bibitem[{{Baluev}(2009)}]{Baluev2009}
{Baluev}, R.~V. 2009, \mnras, 393, 969

\bibitem[{{Beaug{\'e}} {et~al.}(2003){Beaug{\'e}}, {Ferraz-Mello}, \&
  {Michtchenko}}]{Beauge2003}
{Beaug{\'e}}, C., {Ferraz-Mello}, S., \& {Michtchenko}, T.~A. 2003, \apj, 593,
  1124

\bibitem[{{Brahm} {et~al.}(2017{\natexlab{a}}){Brahm}, {Jord{\'a}n}, \&
  {Espinoza}}]{Brahm2017a}
{Brahm}, R., {Jord{\'a}n}, A., \& {Espinoza}, N. 2017{\natexlab{a}}, \pasp,
  129, 034002

\bibitem[{{Brahm} {et~al.}(2017{\natexlab{b}}){Brahm}, {Jord{\'a}n}, {Hartman},
  \& {Bakos}}]{Brahm2017b}
{Brahm}, R., {Jord{\'a}n}, A., {Hartman}, J., \& {Bakos}, G.
  2017{\natexlab{b}}, \mnras, 467, 971

\bibitem[{{Bressan} {et~al.}(2012){Bressan}, {Marigo}, {Girardi}, {Salasnich},
  {Dal Cero}, {Rubele}, \& {Nanni}}]{Bressan2012}
{Bressan}, A., {Marigo}, P., {Girardi}, L., {et~al.} 2012, \mnras, 427, 127

\bibitem[{{Butler} {et~al.}(1996){Butler}, {Marcy}, {Williams}, {McCarthy},
  {Dosanjh}, \& {Vogt}}]{Butler1996}
{Butler}, R.~P., {Marcy}, G.~W., {Williams}, E., {et~al.} 1996, \pasp, 108, 500

\bibitem[{{Butler} {et~al.}(2017){Butler}, {Vogt}, {Laughlin}, {Burt},
  {Rivera}, {Tuomi}, {Teske}, {Arriagada}, {Diaz}, {Holden}, \&
  {Keiser}}]{Butler2017}
{Butler}, R.~P., {Vogt}, S.~S., {Laughlin}, G., {et~al.} 2017, \aj, 153, 208

\bibitem[{{Castelli} \& {Kurucz}(2004)}]{Castelli2004}
{Castelli}, F., \& {Kurucz}, R.~L. 2004, ArXiv Astrophysics e-prints,
  astro-ph/0405087

\bibitem[{{Duncan} {et~al.}(1998){Duncan}, {Levison}, \& {Lee}}]{Duncan1998}
{Duncan}, M.~J., {Levison}, H.~F., \& {Lee}, M.~H. 1998, \aj, 116, 2067

\bibitem[{{ESA}(1997)}]{ESA}
{ESA}, ed. 1997, ESA Special Publication, Vol. 1200, {The HIPPARCOS and TYCHO
  catalogues. Astrometric and photometric star catalogues derived from the ESA
  HIPPARCOS Space Astrometry Mission}

\bibitem[{{Foreman-Mackey} {et~al.}(2013){Foreman-Mackey}, {Hogg}, {Lang}, \&
  {Goodman}}]{emcee}
{Foreman-Mackey}, D., {Hogg}, D.~W., {Lang}, D., \& {Goodman}, J. 2013, PASP,
  125, 306

\bibitem[{{Gaia Collaboration} {et~al.}(2018){Gaia Collaboration}, {Brown},
  {Vallenari}, {Prusti}, {de Bruijne}, {Babusiaux}, \&
  {Bailer-Jones}}]{Gaia_Collaboration2018b}
{Gaia Collaboration}, {Brown}, A.~G.~A., {Vallenari}, A., {et~al.} 2018, ArXiv
  e-prints, arXiv:1804.09365

\bibitem[{{Gaia Collaboration} {et~al.}(2016){Gaia Collaboration}, {Prusti},
  {de Bruijne}, {Brown}, {Vallenari}, {Babusiaux}, {Bailer-Jones}, {Bastian},
  {Biermann}, {Evans}, \& et~al.}]{Gaia_Collaboration2016}
{Gaia Collaboration}, {Prusti}, T., {de Bruijne}, J.~H.~J., {et~al.} 2016,
  \aap, 595, A1

\bibitem[{{Giguere} {et~al.}(2015){Giguere}, {Fischer}, {Payne}, {Brewer},
  {Johnson}, {Howard}, \& {Isaacson}}]{Giguere2015}
{Giguere}, M.~J., {Fischer}, D.~A., {Payne}, M.~J., {et~al.} 2015, \apj, 799,
  89

\bibitem[{{Goldreich} \& {Schlichting}(2014)}]{Goldreich2014}
{Goldreich}, P., \& {Schlichting}, H.~E. 2014, \aj, 147, 32

\bibitem[{{Gontcharov} \& {Mosenkov}(2018)}]{Gontcharov2018}
{Gontcharov}, G.~A., \& {Mosenkov}, A.~V. 2018, VizieR Online Data Catalog,
  2354

\bibitem[{{Ida} {et~al.}(2016){Ida}, {Guillot}, \& {Morbidelli}}]{Ida2016}
{Ida}, S., {Guillot}, T., \& {Morbidelli}, A. 2016, \aap, 591, A72

\bibitem[{{Johnson} {et~al.}(2011){Johnson}, {Payne}, {Howard}, {Clubb},
  {Ford}, {Bowler}, {Henry}, {Fischer}, {Marcy}, {Brewer}, {Schwab}, {Reffert},
  \& {Lowe}}]{Johnson2011}
{Johnson}, J.~A., {Payne}, M., {Howard}, A.~W., {et~al.} 2011, \aj, 141, 16

\bibitem[{{Kjeldsen} \& {Bedding}(2011)}]{Kjeldsen2011}
{Kjeldsen}, H., \& {Bedding}, T.~R. 2011, \aap, 529, L8

\bibitem[{{Lee} \& {Peale}(2002)}]{LeeM2002}
{Lee}, M.~H., \& {Peale}, S.~J. 2002, \apj, 567, 596

\bibitem[{{Lee} \& {Peale}(2003)}]{LeeM2003}
---. 2003, \apj, 592, 1201

\bibitem[{{Lin} \& {Papaloizou}(1986)}]{Lin1986}
{Lin}, D.~N.~C., \& {Papaloizou}, J. 1986, \apj, 309, 846

\bibitem[{{Lubow} \& {D'Angelo}(2006)}]{Lubow2006}
{Lubow}, S.~H., \& {D'Angelo}, G. 2006, \apj, 641, 526

\bibitem[{{Marcy} \& {Butler}(1992)}]{Marcy2}
{Marcy}, G.~W., \& {Butler}, R.~P. 1992, \pasp, 104, 270

\bibitem[{{Marshall} {et~al.}(2018){Marshall}, {Wittenmyer}, {Horner}, {Clark},
  {Mengel}, {Hinse}, {Agnew}, \& {Kane}}]{Marshall2018}
{Marshall}, J.~P., {Wittenmyer}, R.~A., {Horner}, J., {et~al.} 2018, ArXiv
  e-prints, arXiv:1811.06476

\bibitem[{{Masset} \& {Snellgrove}(2001)}]{Masset2001}
{Masset}, F., \& {Snellgrove}, M. 2001, \mnras, 320, L55

\bibitem[{{Masset} \& {Papaloizou}(2003)}]{Masset2003}
{Masset}, F.~S., \& {Papaloizou}, J.~C.~B. 2003, \apj, 588, 494

\bibitem[{{Meschiari} {et~al.}(2009){Meschiari}, {Wolf}, {Rivera}, {Laughlin},
  {Vogt}, \& {Butler}}]{Meschiari2009}
{Meschiari}, S., {Wolf}, A.~S., {Rivera}, E., {et~al.} 2009, \pasp, 121, 1016

\bibitem[{Nelder \& Mead(1965)}]{NelderMead}
Nelder, J.~A., \& Mead, R. 1965, Computer Journal, 7, 308

\bibitem[{{Nelson} {et~al.}(2016){Nelson}, {Robertson}, {Payne}, {Pritchard},
  {Deck}, {Ford}, {Wright}, \& {Isaacson}}]{Nelson2016}
{Nelson}, B.~E., {Robertson}, P.~M., {Payne}, M.~J., {et~al.} 2016, \mnras,
  455, 2484

\bibitem[{{Pierens} {et~al.}(2014){Pierens}, {Raymond}, {Nesvorny}, \&
  {Morbidelli}}]{Pierens2014}
{Pierens}, A., {Raymond}, S.~N., {Nesvorny}, D., \& {Morbidelli}, A. 2014,
  \apjl, 795, L11

\bibitem[{{Pojmanski} {et~al.}(2005){Pojmanski}, {Pilecki}, \&
  {Szczygiel}}]{Pojmanski2005}
{Pojmanski}, G., {Pilecki}, B., \& {Szczygiel}, D. 2005, \actaa, 55, 275

\bibitem[{{Press} {et~al.}(1992){Press}, {Teukolsky}, {Vetterling}, \&
  {Flannery}}]{Press}
{Press}, W.~H., {Teukolsky}, S.~A., {Vetterling}, W.~T., \& {Flannery}, B.~P.
  1992, {Numerical recipes in FORTRAN. The art of scientific computing}
  (Cambridge University Press)

\bibitem[{{Quirrenbach} {et~al.}(2011){Quirrenbach}, {Reffert}, \&
  {Bergmann}}]{Quirrenbach}
{Quirrenbach}, A., {Reffert}, S., \& {Bergmann}, C. 2011, in American Institute
  of Physics Conference Series, Vol. 1331, American Institute of Physics
  Conference Series, ed. S.~{Schuh}, H.~{Drechsel}, \& U.~{Heber}, 102--109

\bibitem[{{Rein} {et~al.}(2012){Rein}, {Payne}, {Veras}, \& {Ford}}]{Rein2012}
{Rein}, H., {Payne}, M.~J., {Veras}, D., \& {Ford}, E.~B. 2012, \mnras, 426,
  187

\bibitem[{{Rivera} {et~al.}(2010){Rivera}, {Laughlin}, {Butler}, {Vogt},
  {Haghighipour}, \& {Meschiari}}]{Rivera2010}
{Rivera}, E.~J., {Laughlin}, G., {Butler}, R.~P., {et~al.} 2010, \apj, 719, 890

\bibitem[{{Robinson} {et~al.}(2007){Robinson}, {Laughlin}, {Vogt}, {Fischer},
  {Butler}, {Marcy}, {Henry}, {Driscoll}, {Takeda}, \&
  {Johnson}}]{Robinson2007}
{Robinson}, S.~E., {Laughlin}, G., {Vogt}, S.~S., {et~al.} 2007, \apj, 670,
  1391

\bibitem[{{Robitaille} {et~al.}(2007){Robitaille}, {Whitney}, {Indebetouw}, \&
  {Wood}}]{Robitaille}
{Robitaille}, T.~P., {Whitney}, B.~A., {Indebetouw}, R., \& {Wood}, K. 2007,
  \apjs, 169, 328

\bibitem[{{Sato} {et~al.}(2012){Sato}, {Omiya}, {Harakawa}, {Izumiura},
  {Kambe}, {Takeda}, {Yoshida}, {Itoh}, {Ando}, {Kokubo}, \& {Ida}}]{Sato2012}
{Sato}, B., {Omiya}, M., {Harakawa}, H., {et~al.} 2012, \pasj, 64, 135

\bibitem[{{Sato} {et~al.}(2016){Sato}, {Wang}, {Liu}, {Zhao}, {Omiya},
  {Harakawa}, {Nagasawa}, {Wittenmyer}, {Butler}, {Song}, {He}, {Zhao},
  {Kambe}, {Noguchi}, {Ando}, {Izumiura}, {Okada}, {Yoshida}, {Takeda}, {Itoh},
  {Kokubo}, \& {Ida}}]{Sato2016}
{Sato}, B., {Wang}, L., {Liu}, Y.-J., {et~al.} 2016, \apj, 819, 59

\bibitem[{{Stock} {et~al.}(2018){Stock}, {Reffert}, \&
  {Quirrenbach}}]{Stock2018}
{Stock}, S., {Reffert}, S., \& {Quirrenbach}, A. 2018, \aap, 616, A33

\bibitem[{{Tadeu dos Santos} {et~al.}(2015){Tadeu dos Santos}, {Correa-Otto},
  {Michtchenko}, \& {Ferraz-Mello}}]{Santos2015}
{Tadeu dos Santos}, M., {Correa-Otto}, J.~A., {Michtchenko}, T.~A., \&
  {Ferraz-Mello}, S. 2015, \aap, 573, A94

\bibitem[{{Tan} {et~al.}(2013){Tan}, {Payne}, {Lee}, {Ford}, {Howard},
  {Johnson}, {Marcy}, \& {Wright}}]{Tan2013}
{Tan}, X., {Payne}, M.~J., {Lee}, M.~H., {et~al.} 2013, \apj, 777, 101

\bibitem[{{Trifonov} {et~al.}(2018){Trifonov}, {Lee}, {Reffert}, \&
  {Quirrenbach}}]{Trifonov2018a}
{Trifonov}, T., {Lee}, M.~H., {Reffert}, S., \& {Quirrenbach}, A. 2018, \aj,
  155, 174

\bibitem[{{Trifonov} {et~al.}(2014){Trifonov}, {Reffert}, {Tan}, {Lee}, \&
  {Quirrenbach}}]{Trifonov2014}
{Trifonov}, T., {Reffert}, S., {Tan}, X., {Lee}, M.~H., \& {Quirrenbach}, A.
  2014, \aap, 568, A64

\bibitem[{{Valenti} {et~al.}(1995){Valenti}, {Butler}, \& {Marcy}}]{Valenti}
{Valenti}, J.~A., {Butler}, R.~P., \& {Marcy}, G.~W. 1995, \pasp, 107, 966

\bibitem[{{VanderPlas} {et~al.}(2012){VanderPlas}, {Connolly}, {Ivezic}, \&
  {Gray}}]{VanderPlas2012}
{VanderPlas}, J., {Connolly}, A.~J., {Ivezic}, Z., \& {Gray}, A. 2012, in
  Proceedings of Conference on Intelligent Data Understanding (CIDU), pp.
  47-54, 2012., 47

\bibitem[{{Vogt} {et~al.}(1994){Vogt}, {Allen}, {Bigelow}, {Bresee}, {Brown},
  {Cantrall}, {Conrad}, {Couture}, {Delaney}, {Epps}, {Hilyard}, {Hilyard},
  {Horn}, {Jern}, {Kanto}, {Keane}, {Kibrick}, {Lewis}, {Osborne},
  {Pardeilhan}, {Pfister}, {Ricketts}, {Robinson}, {Stover}, {Tucker}, {Ward},
  \& {Wei}}]{Vogt1994}
{Vogt}, S.~S., {Allen}, S.~L., {Bigelow}, B.~C., {et~al.} 1994, in \procspie,
  Vol. 2198, Instrumentation in Astronomy VIII, ed. D.~L. {Crawford} \& E.~R.
  {Craine}, 362

\bibitem[{{Walsh} {et~al.}(2011){Walsh}, {Morbidelli}, {Raymond}, {O'Brien}, \&
  {Mandell}}]{Walsh2011}
{Walsh}, K.~J., {Morbidelli}, A., {Raymond}, S.~N., {O'Brien}, D.~P., \&
  {Mandell}, A.~M. 2011, \nat, 475, 206

\bibitem[{{Wittenmyer} {et~al.}(2012){Wittenmyer}, {Horner}, \&
  {Tinney}}]{Wittenmyer2012}
{Wittenmyer}, R.~A., {Horner}, J., \& {Tinney}, C.~G. 2012, \apj, 761, 165

\bibitem[{{Wittenmyer} {et~al.}(2016){Wittenmyer}, {Johnson}, {Butler},
  {Horner}, {Wang}, {Robertson}, {Jones}, {Jenkins}, {Brahm}, {Tinney},
  {Mengel}, \& {Clark}}]{Wittenmyer2016}
{Wittenmyer}, R.~A., {Johnson}, J.~A., {Butler}, R.~P., {et~al.} 2016, \apj,
  818, 35

\bibitem[{{Worthey} \& {Lee}(2011)}]{Worthey}
{Worthey}, G., \& {Lee}, H.-c. 2011, \apjs, 193, 1

\bibitem[{{Zechmeister} \& {K{\"u}rster}(2009)}]{Zechmeister2009}
{Zechmeister}, M., \& {K{\"u}rster}, M. 2009, \aap, 496, 577

\end{thebibliography}





%

\appendix

 \setcounter{table}{0}
\renewcommand{\thetable}{A\arabic{table}}

\setcounter{figure}{0}
\renewcommand{\thefigure}{A\arabic{figure}}

\begin{table*}
\caption{HIRES Doppler measurements for HD\,202696 from \citet{Butler2017} and RV residuals of the Best-fit models from Table~\ref{table:orb_par_stable} } 
\label{table:HD202696} 

\centering  

\begin{tabular}{c r c c c r r r} 

\hline\hline    
\noalign{\vskip 0.5mm}

Epoch (BJD) & RV (m\,s$^{-1}$) & $\sigma_{RV}$ (m\,s$^{-1}$)   &  S index & H index &  o$-$c$_{\rm \small ~Kep.}$  (m\,s$^{-1}$) &  o$-$c$_{\rm \small ~N-body }$  (m\,s$^{-1}$)   \\  

\hline     
\noalign{\vskip 0.5mm}    

2,454,287.927   &   2.88   &    1.96     &  0.1091   &    0.0312  &   $-$2.94  &   $-$3.57  \\ 
2,454,399.822   &   $-$63.90   &    1.25     &  0.1072   &    0.0307  &   $-$2.53  &   $-$4.50  \\ 
2,454,634.038   &   $-$20.49   &    1.14     &  0.0899   &    0.0312  &   $-$5.05  &   $-$5.41  \\ 
2,454,674.900   &   2.57   &    1.39     &  0.1066   &    0.0308  &   $-$0.79  &   $-$0.85  \\ 
2,454,718.006   &   10.64   &    1.49     &  0.0868   &    0.0308  &   $-$4.21  &   $-$4.00  \\ 
2,454,777.857   &   11.66   &    1.19     &  0.1132   &    0.0309  &   $-$4.38  &   $-$3.84  \\ 
2,454,778.819   &   13.33   &    1.39     &  0.1065   &    0.0308  &   $-$2.61  &   $-$2.07  \\ 
2,454,805.740   &   10.92   &    1.40     &  0.1138   &    0.0308  &   $-$1.13  &   $-$0.41  \\ 
2,454,955.107   &   $-$5.77   &    1.32     &  0.1055   &    0.0308  &   16.03  &   16.72  \\ 
2,454,964.128   &   $-$15.69   &    1.23     &  0.1075   &    0.0310  &   7.25  &   7.84  \\ 
2,454,984.066   &   $-$23.92   &    1.38     &  0.1065   &    0.0311  &   0.58  &   0.93  \\ 
2,454,985.007   &   $-$26.07   &    1.40     &  0.1114   &    0.0309  &   $-$1.53  &   $-$1.20  \\ 
2,455,014.963   &   $-$39.57   &    1.31     &  0.1072   &    0.0308  &   $-$15.66  &   $-$15.70  \\ 
2,455,015.957   &   $-$41.36   &    1.31     &  0.1061   &    0.0310  &   $-$17.54  &   $-$17.59  \\ 
2,455,019.021   &   $-$25.42   &    1.30     &  0.1143   &    0.0309  &   $-$1.89  &   $-$1.97  \\ 
2,455,043.881   &   $-$18.51   &    1.40     &  0.1074   &    0.0307  &   1.18  &   0.87  \\ 
2,455,076.033   &   $-$9.97   &    1.37     &  0.1129   &    0.0306  &   1.10  &   0.74  \\ 
2,455,084.036   &   0.00   &    1.61     &  0.1251   &    0.0306  &   8.42  &   8.08  \\ 
2,455,106.909   &   5.94   &    1.47     &  0.1086   &    0.0307  &   6.10  &   5.92  \\ 
2,455,133.918   &   17.65   &    1.40     &  0.0936   &    0.0305  &   7.88  &   7.96  \\ 
2,455,169.772   &   21.46   &    1.40     &  0.1145   &    0.0307  &   1.14  &   1.44  \\ 
2,455,170.691   &   17.93   &    1.43     &  0.1069   &    0.0308  &   $-$2.59  &   $-$2.29  \\ 
2,455,187.695   &   20.55   &    1.39     &  0.1011   &    0.0308  &   $-$2.85  &   $-$2.57  \\ 
2,455,198.700   &   21.99   &    1.42     &  0.1075   &    0.0309  &   $-$2.35  &   $-$2.15  \\ 
2,455,290.146   &   2.47   &    1.33     &  0.1010   &    0.0309  &   0.17  &   $-$1.29  \\ 
2,455,373.126   &   $-$33.29   &    1.23     &  0.1137   &    0.0309  &   7.89  &   7.22  \\ 
2,455,396.059   &   $-$52.38   &    1.36     &  0.1097   &    0.0309  &   $-$1.22  &   $-$1.26  \\ 
2,455,436.743   &   $-$69.28   &    1.27     &  0.0975   &    0.0307  &   $-$5.03  &   $-$4.34  \\ 
2,455,455.739   &   $-$73.42   &    1.34     &  0.1085   &    0.0307  &   $-$5.25  &   $-$4.49  \\ 
2,455,521.793   &   $-$53.00   &    1.37     &  0.1062   &    0.0308  &   16.87  &   16.67  \\ 
2,455,698.124   &   $-$10.58   &    1.35     &  0.1006   &    0.0309  &   $-$16.91  &   $-$17.80  \\ 
2,455,770.042   &   53.71   &    1.56     &  0.1099   &    0.0309  &   28.09  &   27.74  \\ 
2,455,787.946   &   20.04   &    9.66     &  0.0659   &    0.0285  &   $-$6.39  &   $-$6.71  \\ 
2,455,841.869   &   26.89   &    1.26     &  0.0882   &    0.0308  &   5.86  &   5.65  \\ 
2,455,880.836   &   8.95   &    1.23     &  0.1292   &    0.0312  &   $-$3.05  &   $-$3.06  \\ 
2,455,903.717   &   $-$1.19   &    1.25     &  0.1071   &    0.0309  &   $-$6.85  &   $-$6.68  \\ 
2,455,931.693   &   2.89   &    1.29     &  0.0945   &    0.0307  &   5.29  &   5.69  \\ 
2,456,098.123   &   $-$28.24   &    1.31     &  0.1084   &    0.0309  &   $-$8.23  &   $-$7.59  \\ 
2,456,154.011   &   $-$2.81   &    1.46     &  0.1084   &    0.0308  &   4.67  &   5.64  \\ 
2,456,194.986   &   3.05   &    1.31     &  0.1098   &    0.0307  &   2.08  &   3.36  \\ 
2,456,522.092   &   $-$57.44   &    1.43     &  0.1149   &    0.0309  &   1.73  &   0.45  \\ 
2,456,894.081   &   19.32   &    1.23     &  0.1113   &    0.0309  &   $-$5.09  &   $-$5.03  \\ 
  
\hline           
\end{tabular}

\end{table*}

\end{document}